\documentclass[aps,pre,reprint,10pt,floatfix]{revtex4-1}

\usepackage[utf8]{inputenc}

\usepackage[lining,semibold]{libertine} 
\usepackage[libertine, cmintegrals, bigdelims, vvarbb]{newtxmath}

\usepackage{calc}
\usepackage{bm}
\usepackage{microtype}
\usepackage{iftex}
\usepackage[hidelinks]{hyperref}
\usepackage{xcolor}
\usepackage{graphicx}
\usepackage{booktabs}

\usepackage[version=4]{mhchem}

\usepackage{kbordermatrix}
\renewcommand{\kbldelim}{(}
\renewcommand{\kbrdelim}{)}
\definecolor{blue}{rgb}{0, 0.35, 0.60}
\newcommand{\new}[1]{{ {#1} }}
\definecolor{oragne}{rgb}{0.60, 0.35, 0}
\definecolor{green}{rgb}{0, 0.55, 0.30}
\newcommand{\nw}[1]{{{#1}}}

\ifPDFTeX
  
\else
  
\fi

\newcommand{\EP}{\ensuremath{\sigma}}

\newcommand{\stoich}{\ensuremath{\mathbb{S}}}
\newcommand{\diagram}[2]{\vcenter{\hbox{\includegraphics[scale=#2]{#1}}}}

\renewcommand{\vec}[1]{\bm{#1}}

\newcommand{\D}{\mathrm{d}}

\newcommand{\ddt}{\frac{\del}{\del t}}

\newcommand{\del}{\partial}

\newcommand{\subsc}[1]{_{\text{#1}}}

\newcommand{\eq}[1]{(\ref{eq:#1})}
\newcommand{\Eq}[1]{Eq.~\eq{#1}}
\newcommand{\Eqs}[1]{Eqs.~\eq{#1}}
\newcommand{\EQ}[1]{Equation~\eq{#1}}

\newcommand{\fig}[1]{\ref{fig:#1}}
\newcommand{\Fig}[1]{Fig.~\fig{#1}}

\graphicspath{{./figs/}{./diagram/}}

\definecolor{g}{gray}{0.50}

\begin{document}

\hypersetup{pdftex,
            pdfauthor={Artur Wachtel, Riccardo Rao, and Massimiliano Esposito},
            pdftitle={Thermodynamically Consistent Coarse Graining of Biocatalysts beyond Michaelis--Menten},
            pdfproducer={Latex with hyperref},
            pdfcreator={pdflatex}}

\title{Thermodynamically Consistent Coarse Graining of Biocatalysts beyond Michaelis--Menten}

\date{\today}

\author{Artur Wachtel}
\email{artur.wachtel@uni.lu}
\author{Riccardo Rao}
\author{Massimiliano Esposito}
\affiliation{
Complex Systems and Statistical Mechanics, Physics and Materials Science Research Unit, University of Luxembourg, 162a, Avenue de la Fa\"{i}encerie, 1511 Luxembourg, G. D. Luxembourg
}
\begin{abstract}
  Starting from the detailed catalytic mechanism of a biocatalyst we provide a coarse-graining procedure which, by construction, is thermodynamically consistent.
  This procedure provides stoichiometries, reaction fluxes (rate laws), and reaction forces (Gibbs energies of reaction) for the coarse-grained level.
  It can treat active transporters and molecular machines, and thus extends the applicability of ideas that originated in enzyme kinetics.
  Our results lay the foundations for systematic studies of the thermodynamics of large-scale biochemical reaction networks.
  Moreover, we identify the conditions under which a relation between one-way fluxes and forces holds at the coarse-grained level as it holds at the detailed level.
  In doing so, we clarify the speculations and broad claims made in the literature about such a general flux--force relation.
  As a further consequence we show that, in contrast to common belief, the second law of thermodynamics does not require the currents and the forces of biochemical reaction networks to be always aligned.
\end{abstract}
\maketitle


\section{Introduction}

Catalytic processes are ubiquitous in cellular physiology.
%
Biocatalysts are involved in metabolism, cell signalling, transcription and translation of genetic information, as well as replication.
All these processes and pathways involve not only a few but rather dozens to hundreds, sometimes thousands of different enzymes. 
Finding the actual catalytic mechanism of a single enzyme is difficult and time consuming work.
To date, for many enzymes the catalytic mechanisms are not known.
Even if such detailed information was at hand, including detailed catalytic machanisms into a large scale model is typically unfeasable for numerical simulations.
Therefore, larger biochemical reaction networks contain the enzymes as single reactions following enzymatic kinetics.
This simplified description captures only the essential dynamical features of the catalytic action, condensed into a single reaction.

The history of enzyme kinetics~\cite{Cornish-Bowden2013} stretches back more than a hundred years.
After the pioneering work of Brown~\cite{Brown1902} and Henri~\cite{Henri1902}, Michaelis and Menten~\cite{Michaelis.Menten1913} layed the foundation for the systematic coarse graining of a detailed enzymatic mechanism into a single reaction.
Since then, a lot of different types of mechanisms have been found and systematically classified~\cite{Cornish-Bowden2012}.
Arguably, the most important catalysts in biochemical processes are enzymes -- which are catalytically active proteins.
However, other \nw{types of} catalytic molecules are \nw{also} known, some of them occur naturally like catalytic RNA (ribozymes) or catalytic anti-bodies (abzymes), some of them are synthetic (synzymes) \cite{Cornish-Bowden2012}.
For our purposes it does not matter which kind of biocatalyst is being described by a catalytic mechanism -- we treat all of the above in the same way.

From a more general perspective, other scientific fields are concerned with the question of how to properly coarse grain a process \nw{as well}.
While in most applications the focus lies on the dynamics, or kinetics, of a process, it turned out that thermodynamics plays an intricate role in this question~\cite{Puglisi.etal2010}.
\new{For processes occurring at thermodynamic equilibrium, every choice of coarse graining can be made thermodynamically consistent}  -- after all, the very foundation of equilibrium thermodynamics is concerned with reduced descriptions of physical phenomena~\cite{Callen1985}.
Instead, biological systems are open systems exchanging particles with reservoirs and as such they are inherently out of equilibrium.
Nonequilibrium processes, in general, do not have a natural coarse graining.

\new{
When the particle numbers in a reaction network are small, it needs to be described stochastically with the chemical master equation.
Indeed, there is increased interest in the correct thermodynamic treatment of stochastic processes~\cite{Seifert2012, VandenBroeck.Esposito2015}.
With stochastic processes it is possible to investigate energy-conversion in molecular motors~\cite{Juelicher.etal1997,Seifert2011a, Altaner.etal2015, Astumian.etal2016}, error correction via kinetic proofreading~\cite{Murugan.etal2012, Rao.Peliti2015, Sartori.Pigolotti2015}, as well as information processing in small sensing networks~\cite{Barato.etal2014, Bo.Celani2016, Ouldridge.etal2017}.
In this field, different suggestions arose for coarse grainings motivated by thermodynamic consistency~\cite{Esposito2012, Altaner.Vollmer2012, Knoch.Speck2015}.
In these cases, the dissipation in a nonequililibrium process is typically underestimated -- although also overestimations may occur~\cite{Esposito.Parrondo2015}.
For a general overview of coarse-graining in Markov processes, see Ref.~\cite{Bo.Celani2017} and references therein.

For large-scale networks however, a stochastic treatment is unfeasable.
On the one hand, \nw{stochastic simulations quickly become computationally so demanding that they are effectively} untractable.
On the other hand, when species appear in large abundances (e.g. metabolic networks) the stochastic noise is negligible.
This paper is exclusively concerned with this latter case.
The dynamics is governed by deterministic differential equations: the non-linear rate equations of chemical kinetics.
Assuming a separation of time scales in these equations, model reduction approaches have been developed~\cite{Segel.Slemrod1989, Gunawardena2014, Rubin.etal2014}.
However, they do not address the question of thermodynamic consistency.
Remarkably, recent development in the thermodynamics of chemical reaction networks~\cite{Polettini.Esposito2014, Rao.Esposito2016} highlighted the strong connection between the thermodynamics of deterministic rate equations and of stochastic processes, including the relation between energy, work, and information.
Unfortunately, these studies were limited to elementary reactions with mass--action kinetics.
The present paper addresses this constraint, thus extending the theory to kinetics of coarse-grained catalysts.}

\begin{figure*}[t]
  \centering
    \includegraphics[width=0.95\columnwidth]{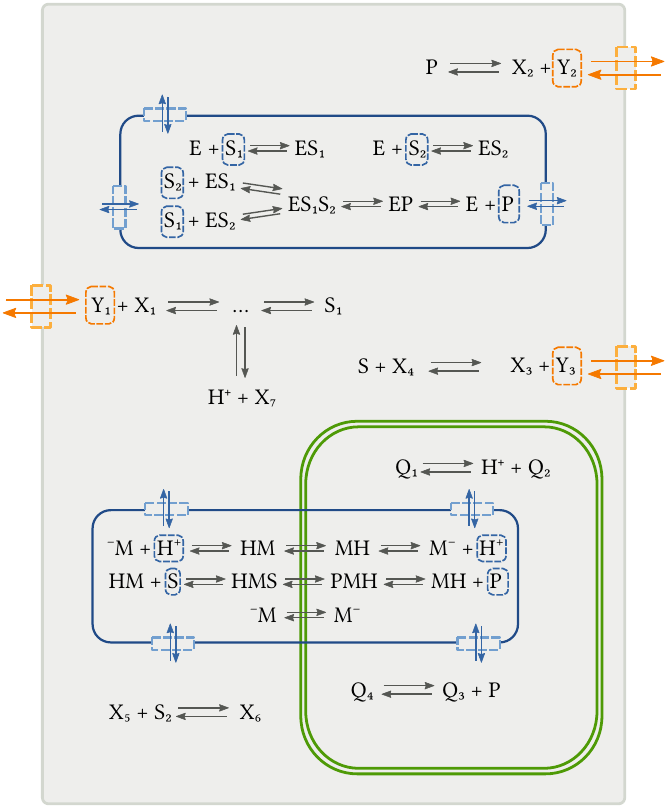}
  \hspace{2em}
    \includegraphics[width=0.95\columnwidth]{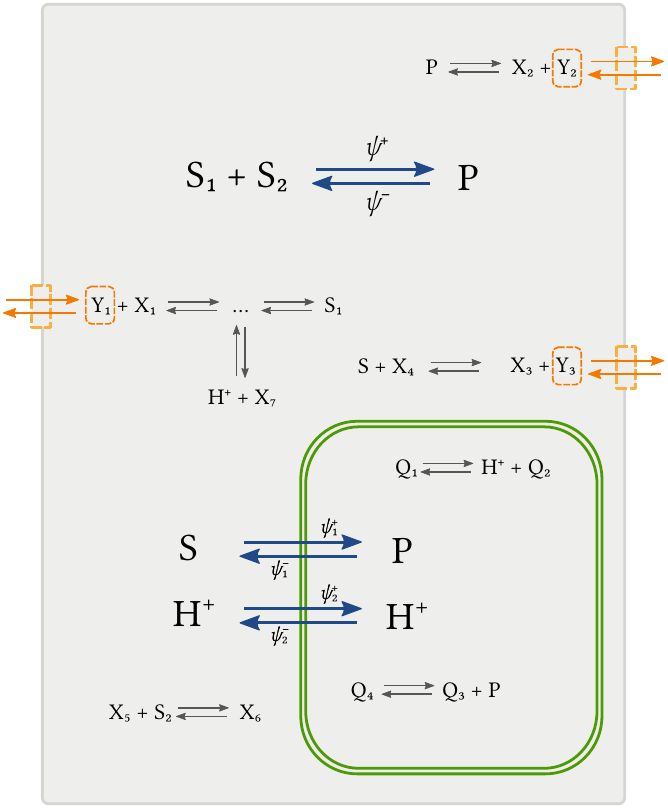}
\caption{\new{
  Overview of the coarse-graining procedure:
  [left] The starting point is a reaction network with elementary reactions following mass--action kinetics in a steady state.
  This example contains two catalytic mechanisms (blue boxes) \nw{and for illustrative purposes some additional arbitrary reactions}.  Each of the two catalyst species, \(\ce{E}\) and \(\ce{M}\), is conserved throughout the network.  The reaction partners of the catalysts re-appear in the rest of the network.  From the perspective of the remaining network, only the turnover (blue arrows) of these molecules are relevant.  The involved concentrations may be global, as for \(\ce{S}\), or refer to different well stirred sub-compartments (green box), as for \(\ce{P}\).
  [right]
  The procedure provides few coarse-grained reactions (blue arrows) that replace the originally more complicated mechanisms.  The kinetic rate laws, \(\psi\), of the coarse-grained reactions are different from mass--action.  We construct them explicitly during the coarse-graining procedure, so that the turnover is correctly reproduced.  
  Combined with the coarse-grained reaction forces (Gibbs free energies) also the entropy-production rate is reproduced exactly.
  We work out the coarse graining of these two catalysts, \(\ce{E}\) and \(\ce{M}\), in detail in section~\ref{sec:examples}.
}}
  \label{fig:cg}
\end{figure*}

Understanding the nonequilibrium thermodynamics of catalysts is a crucial step towards incorporating thermodynamics into large-scale reaction networks.
There is ongoing effort in the latter~\cite{Beard.etal2002, Fleming.etal2010, Soh.Hatzimanikatis2010}
which very often is based on the connection between thermodynamics and kinetics~\cite{Beard.Qian2007, Wiechert2007, Noor.etal2013}.

In this paper we show how to coarse grain the \new{deterministic} description of any biocatalyst in a thermodynamically consistent way -- extending the applicability of such simplifications even to molecular motors~\cite{Juelicher.etal1997, Liepelt.Lipowsky2007} and active membrane transport~\cite{Pietrobon.Caplan1985}. 
The starting point is the catalytic mechanism described as a reversible chemical reaction network where each \new{of the \(M\) reaction steps} \(\rho\) is an elementary \nw{transition representing a conformational change of a molecule or an elementary chemical} reaction with mass--action \nw{kinetics.
  The corresponding rates are given by the fluxes (kinetic rate laws), \(\phi^{\pm}_{\rho}\), that incorporate the reaction rate constants and the dependence on the concentration of the reactant molecules.
The mass--action} reaction forces (negative Gibbs free energies of reaction) are \(-\Delta_{\rho}G = RT \ln \phi^+_{\rho} / \phi^{-}_{\rho}\)~\cite{Alberty2003}.
At this level, the reaction currents, \(J_{\rho} = \phi_{\rho}^{+} - \phi_{\rho}^{-}\), of these elementary steps are aligned with their respective reaction forces~\citep{Kondepudi.Prigogine2015}\nw{: when one is positive, so is the other.}
From here we construct a reduced set of \(C\) reactions with effective reaction fluxes \(\psi^{\pm}_{\alpha}\) and net forces \(-\Delta_{\alpha} G\).
\nw{As we will see later, there is a limited freedom to choose the exact set of reduced reactions.
Nonetheless, the number of reduced reactions is independent of this choice.}

By construction, our coarse graining procedure captures the entropy-production rate (EPR)~\cite{Groot.Mazur1984, Kondepudi.Prigogine2015} of the underlying catalytic mechanism,
\begin{align*}
  T \sigma := - \sum_{\rho}^{\new{M}} \left( \phi^+_{\rho} - \phi^{-}_{\rho} \right) \Delta_{\rho} G = - \sum_{\alpha}^{\new{C}} \left( \psi^{+}_{\alpha} - \psi^{-}_{\alpha} \right) \Delta_{\alpha} G \geq 0\,,
\end{align*}
\new{even though the number \(C\) of effective reactions \(\alpha\) is much smaller than the number \(M\) of original reaction steps \(\rho\).
Therefore, our procedure} is applicable in nonequilibrium situations, such as biological systems.
In fact, the above equation is exact under steady-state conditions.
In transient and other time-dependent situations the coarse graining \new{can be} a reasonable approximation.
We elaborate this point further in the \new{discussion}.

Secondly, we work out the condition for this coarse graining to reduce to a single reaction \(\alpha\).
In this case, we prove that the following flux--force relation holds true for this coarse-grained reaction:
\begin{align*}
  - \Delta_{\alpha} G = RT \ln \frac{\psi^+_{\alpha}}{\psi^-_{\alpha}} \,.
\end{align*}
 A trivial consequence is that the coarse-grained reaction current, \(J_{\alpha} = \psi^+_{\alpha} - \psi^-_{\alpha}\), is aligned with the net force, \(-\Delta_{\alpha} G\).
In the past, such a flux--force relation has been used in the literature~\cite{Flamholz.etal2013, Noor.etal2014} after its general validity was claimed~\cite{Beard.Qian2007} and later questioned~\cite{Wiechert2007, Fleming.etal2010}.
From here the belief arises that in every biochemical reaction network with any type of kinetics the currents and the forces of each reaction individually need to be aligned\new{, a constraint used especially in flux balance analysis~\cite{Henry.etal2007, Orth.etal2010, Chakrabarti.etal2013}}.
\new{However, as we show in this paper, this relation does not hold} when the coarse-graining reduces the biocatalyst to two or more coupled reactions.


This paper is structured as follows: First we present our results.
Then, we illustrate our findings with two examples: The first is enzymatic catalysis of two substrates into one product.
This can be reduced to a single reaction, for which we verify the flux--force relation at the coarse-grained level.
The second example is a model of active membrane transport of protons, which is a prototype of a biocatalyst that cannot be reduced to a single reaction.
\new{Afterwards, we sketch the proofs for our general claims.
Finally, we discuss our results and their implications.}
Rigorous proofs are provided in the \new{appendix}.

\section{Results}

Our main result is a systematic procedure for a thermodynamically consistent coarse graining of catalytic processes.
These processes may involve several substrates, products, modifiers (e.g. activators, inhibitors) that bind to or are released from a single molecule -- the catalyst.
The coarse graining involves only a few steps\new{ and is exemplified graphically in Fig.~\ref{fig:cg}}:

(1) Consider the catalytic mechanism in a closed box and identify the internal stoichiometric cycles of the system.
An internal stoichiometric cycle is a sequence of reactions leaving the state of the system invariant.
Formally, internal stoichiometric cycles constitute the nullspace of the full stoichiometric matrix, \(\stoich\).

(2) Consider the concentrations of all substrates, modifiers, and products (summarized as \(Y\)) constant in time -- therefore reduce the stoichiometric matrix by exactly those species.
\new{The remaining species, \(X\), represent \(N\) different states of the catalyst.}
As a consequence, the reduced stoichiometric matrix, \(\stoich^{X}\), has a larger nullspace: New stoichiometric cycles emerge in the system.
\nw{These emergent cycles cause a turnover in the substrates/products while leaving the internal species invariant.}
\new{Choose a} \nw{basis, \(\vec{C}_\alpha\), of emergent stoichiometric cycles that are linearly independent from the internal cycles.}

(3) Identify the net stoichiometry, \(\stoich^{Y}\vec{C}_\alpha\), together with the sum, \(-\Delta_\alpha G\), of the forces along each \nw{emergent} cycle \(\alpha\).

(4) Calculate the apparent fluxes \(\psi^\pm_\alpha\) along the emergent cycles at steady state. 

\new{
  For each emergent stoichiometric cycle \(\alpha\) this procedure provides a new reversible reaction with net stoichiometry \( \stoich^{Y}\vec{C}_\alpha\), net force \(-\Delta_\alpha G\), and net fluxes \(\psi^\pm_\alpha\).
Furthermore, it preserves the EPR and, therefore, is thermodynamically consistent.
}

Our second result is a consequence of the main result: We prove that the flux--force relation is satisified at the coarse-grained level by any catalytic mechanism for which only one single cycle emerges in step 2 of the presented procedure\new{, as in example~\ref{sec:enzyme}.
When more cycles emerge, the flux--force relation does not hold as we show in the explicit counter-example~\ref{sec:transporter}}.

\section{Examples}
\label{sec:examples}

\subsection{Enzymatic catalysis}
\label{sec:enzyme}



Let us consider \new{the enzyme \ce{E} that we introduced in Fig.~\ref{fig:cg}.  It} is capable of catalyzing a reaction of two substrates, \ce{S1} and \ce{S2}, into a single product molecule, \ce{P}.
\new{The} binding order of the two substrates does not matter.
Every single one of these reaction steps is assumed to be
reversible and to follow mass--action kinetics.
For every reaction we adopt a reference forward direction.
Overall, the enzymatic catalysis can be represented by the reaction network in \Fig{enzyme1}.

We apply our main result to this enzymatic scheme and thus construct a coarse-grained description for the net catalytic action.
We furthermore explicitly verify our second result by showing that the derived enzymatic reaction fluxes satisfy the flux--force relation.

\subsubsection*{(1) Closed system -- internal cycles}

When this system is contained in a closed box, no molecule can leave or enter the reaction volume.
The dynamics is then described by the following rate equations:
\begin{align}
  \frac{\D}{\D t}\vec{z} = \stoich\, \vec{J}(\vec{z})\,,
  \label{eq:rate-eq}
\end{align}
where we introduced the concentration vector and the current vector,
\begin{align*}
  \vec{z} &= 
  \begin{pmatrix}
    [\ce{E}]	\\
    [\ce{ES1}] \\
    [\ce{ES2}] \\
    [\ce{ES1S2}] \\
    [\ce{EP}]	\\
    [\ce{S1}]	\\
    [\ce{S2}]	\\
    [\ce{P}]	
  \end{pmatrix}
  \,,
 &
  \vec{J}(\vec{z}) &=
  \begin{pmatrix}
    k_{1}[\ce{E}][\ce{S1}]  - k_{-1}[\ce{ES1}] \\
    k_{2}[\ce{E}] [\ce{S2}]  - k_{-2}[\ce{ES2}] \\
    k_{3}[\ce{ES2}] [\ce{S1}] - k_{-3}[\ce{ES1S2}] \\
    k_{4}[\ce{ES1}] [\ce{S2}]  - k_{-4}[\ce{ES1S2}] \\
    k_{5}[\ce{ES1S2}] - k_{-5}[\ce{EP}]  \\
    k_{6}[\ce{EP}] - k_{-6}[\ce{E}] [\ce{P}]  
  \end{pmatrix}
  \,,
\end{align*}
as well as the stoichiometric matrix\,,
\begin{align}
  \stoich = 
      \begin{pmatrix}
	-1 & -1 & 0 & 0 & 0 & 1\\
	1 & 0 & 0 & -1 & 0 & 0 \\
	0 & 1 & -1 & 0 & 0 & 0 \\
	0 & 0 & 1 & 1 & -1 & 0 \\
	0 & 0 & 0 & 0 & 1 & -1 \\
	-1 & 0 & -1 & 0 & 0 & 0\\
	0 & -1 & 0 & -1 & 0 & 0\\
	0 & 0 & 0 & 0 & 0 & 1     
      \end{pmatrix}
      \,.
  \label{eq:expl1-stoich}
\end{align}
%

\begin{figure}[t]
  \centering
  \includegraphics{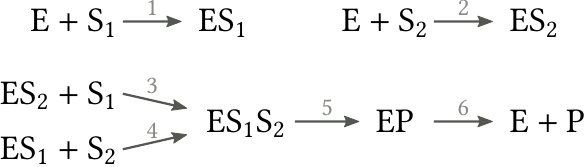}
  \caption{
    An enzymatic scheme turning two substrates into one product.
    The substrates can bind in arbitrary order.
    We adopt a reference direction for the indivinual reactions: forward is from left to right, as indicated by the arrows.
    The backward reactions are from right to left, thus every single reaction step is
    reversible.
    \new{
    This scheme has a clear interpretation as a graph: The reactions are edges, reactants/products are vertices, where different combinations of reactants/products are considered different vertices.
    This graph has three disconnected components and contains no circuit.
    }
  }
  \label{fig:enzyme1}
\end{figure}

In the dynamical equations, only the currents \(\vec{J}(\vec{z})\) depend on the concentrations, whereas the stoichiometric matrix \(\stoich\) does not.
The stoichiometric matrix thus imposes constraints on the possible steady-state concentrations that can be analyzed with mere stoichiometry:
At steady state the current has to satisfy \(\vec{0} = \stoich\,\vec{J}(\vec{z}\subsc{ss})\) or, equivalently, \(\vec{J}(\vec{z}\subsc{ss}) \in \ker \stoich\).
In our example, the null-space \(\ker \stoich\) is one-dimensional and spanned by
\(  \vec{C}\subsc{int} = 
  \begin{pmatrix}
    1 & -1 & -1 & 1 & 0 & 0      
  \end{pmatrix}^{\intercal}\,.\)
  Hence, the steady-state current is fully described by a single scalar value,
  \( \vec{J}(\vec{z}\subsc{ss}) = J\subsc{int} \,\vec{C}\subsc{int}\,.\)
  The vector \(\vec{C}\subsc{int}\) represents a series of reactions that leave the system state unchanged: The two substrates are bound along reactions \(1\) and \(4\) and released again along reactions \(-3\) and \(-2\).
  In the end, the system returns to the exact same state as before these reactions.
Therefore, we call this vector \emph{internal stoichiometric cycle}.
Having identified this internal cycle renders the first step complete.

\new{Note that this stoichiometric cycle does not correspond to a self-avoiding closed path, or \emph{circuit}, in the reaction graph in \Fig{enzyme1}.
This is due to the fact that \emph{combinations} of species serve as vertices.
If instead each species individually is a vertex, then also each cycle corresponds to a circuit.}

In the following we explain why \new{the first} step of the procedure is important.
The closed system has to satisfy a constraint that comes from physics:
A closed system necessarily has to relax to a thermodynamic equilibrium state -- which is characterized by the absence of currents of extensive quantities on any scale.
Thus thermodynamic equilibrium is satisfied if \(J\subsc{int}= 0\).
One can show that this requirement is equivalent to Wegscheider's condition~\cite{Schuster.Schuster1989}: The product of the forward rate constants along the internal cycle equals that of the backward rate constants,
\begin{align}
  k_{1}k_{4}k_{-3}k_{-2} = k_{-1}k_{-4}k_{3}k_{2} \,.
  \label{eq:expl:zero-force}
\end{align}

Furthermore, irrespective of thermodynamic equilibrium, the steady state has to be stoichiometrically compatible with the initial condition:
There are three linearly independent vectors in the cokernel of \(\stoich\):
\vspace{-1em}
\begin{align*}
  \vec{\ell}_{\ce{E}} &= \kbordermatrix{
    & \cr
    \color{g}\ce{E} 	&1 \cr
    \color{g}\ce{ES1} 	&1 \cr
    \color{g}\ce{ES2} 	&1 \cr
    \color{g}\ce{ES1S2}	&1 \cr
    \color{g}\ce{EP} 	&1 \cr
    \color{g}\ce{S1} 	&0 \cr
    \color{g}\ce{S2} 	&0 \cr
    \color{g}\ce{P}  	&0 \cr
  }\,,
  &
  \vec{\ell}_{1} &= \kbordermatrix{
    & \cr
    \color{g}\ce{E} 	&0 \cr
    \color{g}\ce{ES1} 	&1 \cr
    \color{g}\ce{ES2} 	&0 \cr
    \color{g}\ce{ES1S2}	&1 \cr
    \color{g}\ce{EP} 	&1 \cr
    \color{g}\ce{S1} 	&1 \cr
    \color{g}\ce{S2} 	&0 \cr
    \color{g}\ce{P}  	&1 \cr
  }\,,
  &
  \vec{\ell}_{2} &= \kbordermatrix{
    & \cr
    \color{g}\ce{E} 	&0 \cr
    \color{g}\ce{ES1} 	&0 \cr
    \color{g}\ce{ES2} 	&1 \cr
    \color{g}\ce{ES1S2}	&1 \cr
    \color{g}\ce{EP} 	&1 \cr
    \color{g}\ce{S1} 	&0 \cr
    \color{g}\ce{S2} 	&1 \cr
    \color{g}\ce{P}  	&1 \cr
  }\,.
\end{align*}
For each such vector, the scalar \(L \equiv \vec{\ell} \cdot \vec{z}\) evolves according to \(\ddt \vec{\ell} \cdot \vec{z} = \vec{\ell}\cdot \stoich\,\vec{J}(\vec{z}) = 0\), and thus is a \emph{conserved quantity}.
We deliberately chose linearly independent vectors with a clear physical interpretation.
These vectors represent conserved moieties, i.e. a part of (or an entire) molecule that remains intact in all reactions.
The total concentration of the enzyme moiety in the system is given by \(L_{\ce{E}}\).
The other two conservation laws, \(L_1\) and \(L_2\), are the total concentrations of moieties of the substrates, \(\ce{S1}\) and \(\ce{S2}\), respectively.

Given a set of values for the conserved quantities from the initial condition, Wegscheider's condition on the rate constants ensures uniqueness of the equilibrium solution~\cite{Schuster.Schuster1989}.

\subsubsection*{(2) Open system -- emergent cycles}

So far we only discussed the system in a closed box that will necessarily relax to a thermodynamic equilibrium.

\begin{figure}[t]
  \centering
  \includegraphics{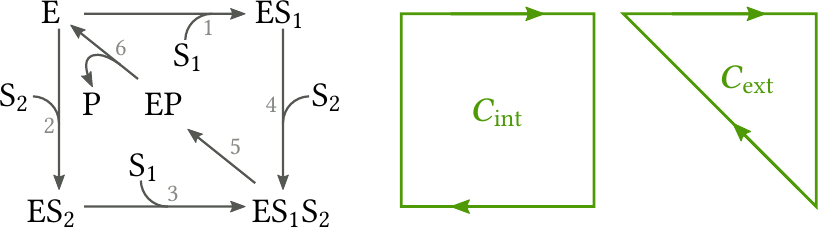}
  \caption{
    [left] Enzymatic catalysis as an \emph{open} chemical network.
    The species \(\ce{S1}\), \(\ce{S2}\) and \(\ce{P}\) are now associated to the edges of the graph, instead of being part of its vertices as in \Fig{enzyme1}.
    \new{This graph has only one connected component and contains three distinct circuits.}
    [center, right] Graphical representation of the two circuits spanning the kernel of \(\stoich^{X}\). \new{The lower left triangle constitutes the third circuit.
    It can be recovered by a linear combination of the other two circuits.}
  } 
  \label{fig:expl-open}
\end{figure}

We now open the box and assume that there is a mechanism capable of fixing the concentrations of \(\ce{S1}\), \(\ce{S2}\) and \(\ce{P}\) to some given levels.
These three species therefore no longer take part in the dynamics.
Formally, we divide the set of species into two disjoint sets: 
\begin{align*}
  \underbrace{\left\{ \ce{E}, \ce{ES1}, \ce{ES2}, \ce{ES1S2}, \ce{EP}\right\}}_{X} \cup \underbrace{\left\{\ce{S1}, \ce{S2}, \ce{P}\right\}}_{Y}\,.
\end{align*}
The first are the \emph{internal species}, \(X\), which are subject to the dynamics.
The second are the \emph{chemostatted species}, \(Y\), which are exchanged with the environment.
We apply this splitting to the stoichiometric matrix,
\begin{align*}
  \stoich &= \begin{pmatrix} \stoich^{X} \\ \stoich^{Y}\end{pmatrix}\,,
\end{align*}
and the vector of concentrations, \(\vec{z} = \left( \vec{x}, \vec{y} \right) \).
Analogously, the rate equations for this open reaction system split into
\begin{align}
  \frac{\del}{\del t}\vec{x} &= \stoich^{X} \; \vec{J}\left( \vec{x}, \vec{y} \right)\,,
  \label{eq:rate-equations--internal}\\
  0 \equiv \frac{\del}{\del t} \vec{y} &= \stoich^{Y} \; \vec{J}\left( \vec{x}, \vec{y} \right) + \vec{I}\left( \vec{x},\vec{y} \right)\,.
  \label{eq:rate-equations--external}
\end{align}
The \EQ{rate-equations--external} is merely a definition for the \new{exchange} current
\(\vec{I}\), keeping the species \(Y\) at constant concentrations.
\new{Note that the exchange currents \(\vec{I}\) quantify the substrate/product turnover.}
The actual dynamical rate equations, the \Eqs{rate-equations--internal}, are a subset of the original equations for the closed system, treating the chemostats as constant parameters.
Absorbing these latter concentrations into the rate constants, we arrive at a linear ODE system with new \emph{pseudo-first-order rate constants} \(\tilde{k}(\vec{y})\).
For these rate equations, one needs to reconsider the graphical representation of this reaction network:
Since the chemostatted species now are merely parameters for the reactions, we have to remove the chemostatted species from the former vertices of the network representation and associate them to the edges.
The resulting graph representing the open network is drawn in \Fig{expl-open}.


The steady-state current \(\vec{J}\subsc{ss} = \vec{J}(\vec{x}\subsc{ss},\vec{y})\) of \Eq{rate-equations--internal} needs to be in the kernel of the internal stoichiometric matrix \(\stoich^{X}\) only.
This opens up new possibilities.
It is obvious that \(\ker \stoich\) is a subset of \(\ker \stoich^{X}\), but \(\ker \stoich^{X}\) is in fact bigger.
In our example we now have two \emph{stoichiometric} cycles,
\vspace{-1em}
\begin{align}
  \vec{C}\subsc{int} &= 
  \kbordermatrix{
    & \cr
    \color{g}1& 1\cr
    \color{g}2& -1\cr  
    \color{g}3& -1\cr  
    \color{g}4& 1\cr    
    \color{g}5& 0\cr   
    \color{g}6& 0      
  }
  & &\text{and} &
  \vec{C}\subsc{ext} &= 
  \kbordermatrix{
    & \cr
    \color{g}1& 1\cr
    \color{g}2& 0\cr  
    \color{g}3& 0\cr  
    \color{g}4& 1\cr    
    \color{g}5& 1\cr   
    \color{g}6& 1      
  }\,.
  \label{eq:expl:stoich-cycles}
\end{align}
The first cycle is the internal cycle we identified in the closed system already: \new{It} only involves reactions that leave the closed system invariant, thus upon completion of this cycle not a single molecule is being exchanged.
The second cycle is different: Upon completion it leaves the internal species unchanged but chemostatted species are exchanged with the environment.
Since this type of cycle appears only upon chemostatting, we call them \emph{emergent} stoichiometric cycles.
Overall, the steady-state current is a linear combination of these two cycles:
\(
  \vec{J}\subsc{ss} = J\subsc{int}\,\vec{C}\subsc{int} + J\subsc{ext} \, \vec{C}\subsc{ext}
  \).
  This completes step 2.

\new{These two stoichiometric cycles correspond to circuits in the open reaction graph. We give a visual representation on the right of \Fig{expl-open}.
  As a consequence of working with catalysts, the vertices of the reaction graph for the open system coincide with the internal species \(X\).
Therefore, for all catalysts the cycles of the open system correspond to circuits in the corresponding graph.}

The cycles are not the only structural object affected by the chemostatting procedure: The conservation laws change as well.
In the enzyme example we have merely one conservation law left -- that of the enzyme moiety, \(L_{\ce{E}}\).
The substrate moieties are being exchanged with the environment, which renders \(L_{1}\) and \(L_{2}\) \emph{broken conservation laws}.
Overall, upon adding three chemostats two conservation laws were broken and one cycle emerged.
In fact, the number of chemostatted species always equals the number of broken conservation laws plus the number of emergent cycles \cite{Polettini.etal2015a}.

\subsubsection*{(3) Net stoichiometries and net forces}

The net stoichiometry of the emergent cycle is \(\ce{S1} + \ce{S2} \rightleftharpoons P \)\,.
This represents a single
reversible reaction describing the net catalytic action of the enzyme.
For a complete coarse graining, we still need to identify the fluxes and the net force along this reaction.
Its net force is given by the sum of the forces along the emergent cycle.
\new{Collecting the Gibbs energies of reaction in a vector, \(\Delta\subsc{r}\vec{G}\coloneqq(\Delta_{1} G, \dots, \Delta_{6} G)^{\intercal}\), this sum is concisely written as}
\begin{align}
  -\Delta\subsc{ext}G := - \vec{C}\subsc{ext} \cdot \Delta\subsc{r}\vec{G} =  RT \ln \frac{k_{1}k_{4}k_{5}k_{6}[\ce{S1}][\ce{S2}]}{k_{-1}k_{-4}k_{-5}k_{-6}[\ce{P}]}\,.
  \label{eq:net-delta-G}
\end{align}

One could also ask about the net stoichiometry and net force along the internal cycle.
However, \new{we have \(\stoich\;\vec{C}\subsc{int}=0\) since the internal cycle does not interact with the chemostats.}
Moreover, the net force along the internal cycle is
\begin{align}
  -\vec{C}\subsc{int} \cdot \Delta\subsc{r}\vec{G} = RT \ln \frac{k_{1}k_{4}k_{-3}k_{-2}}{k_{-1}k_{-4}k_{3}k_{2}} = 0 \label{eq:expl--wegscheider}
\end{align}
by virtue of Wegscheider's condition.

\subsubsection*{(4) Apparent fluxes}

We now determine the apparent fluxes along the two cycles of the system.
To that end, we first \nw{solve the linear rate equations to} calculate the steady-state concentrations and the steady-state currents.
\new{
For the steady-state concentrations we use a diagrammatic method popularized by King and Altman~\cite{King.Altman1956}
\nw{that we summarize} in appendix~\ref{sec:appendix--steady-state}.
}

\nw{
  As derived in step 2 of the procedure, the steady-state current vector is
  \vspace{-1em}
  \begin{align*}
    \vec{J}\subsc{ss} &= 
    \kbordermatrix{
      & \cr
      \color{g}1& J\subsc{int} + J\subsc{ext}\cr
      \color{g}2& -J\subsc{int}\cr  
      \color{g}3& -J\subsc{int}\cr  
      \color{g}4& J\subsc{int} + J\subsc{ext}\cr    
      \color{g}5& J\subsc{ext}\cr   
      \color{g}6& J\subsc{ext}      
    }\,.
  \end{align*}
  Hence the two cycle currents are
  \begin{align*}
    J\subsc{int} &= -J_{2} = k_{-2} [\ce{ES2}] - k_2 [\ce{E}][\ce{S2}]\,, \\
    J\subsc{ext} &= J_{6} = k_6 [\ce{EP}] - k_{-6} [\ce{E}] [\ce{P}]\,.
  \end{align*}
  With the explicit steady-state concentrations given in appendix~\ref{sec:appendix--steady-state--enzyme}, we
  find (see appendix~\ref{sec:appendix--ratelaws--enzyme} for details):
  \begin{align*}
    J\subsc{int} &= k_{-2} [\ce{ES2}] - k_2 [\ce{E}][\ce{S2}] \\
    &= \frac{L_{\ce{E}}k_{2}k_{3}}{N_{\ce{E}}(\vec{y}) k_{1}}\! \left( \frac{k_{-1}}{k_{4}} +  [\ce{S2}] \right)\! \left(  k_{-1}k_{-4}k_{-5}k_{-6} [\ce{P}] - k_{1}k_{4}k_{5}k_{6} [\ce{S1}] [\ce{S2}] \right)\,.
  \end{align*}
  and
\begin{align}
  J\subsc{ext} &= k_6 [\ce{EP}] - k_{-6} [\ce{E}] [\ce{P}] \nonumber \\
  &= \frac{L_{\ce{E}}\xi(\vec{y})}{N_{\ce{E}}(\vec{y})} \left(  k_{1}k_{4}k_{5}k_{6} [\ce{S1}] [\ce{S2}] - k_{-1}k_{-4}k_{-5}k_{-6} [\ce{P}]\right)\,.
  \label{eq:expl--open-steady-current}
\end{align}
Here, \(L_{\ce{E}}\) is the total amount of available enzyme, \(N_{\ce{E}}(\vec{y})\) is a positive quantity that depends on the chemostat concentrations as well as all rate constants, and \[\xi(\vec{y}) = k_{3}[\ce{S1}] + \frac{k_{2}k_{3}[\ce{S2}]}{k_{1}} + k_{-2} + \frac{k_{-2}k_{-3}}{k_{-4}}\,.\]
}

As expected, the current along the emergent cycle \(J\subsc{ext}\) is not zero, provided that its net force is not zero.
However, note that the current along the internal cycle does \emph{not} vanish either, even though its own net force is zero.
Both currents vanish only when the net force, \(-\Delta\subsc{ext}G\), vanishes -- which is at thermodynamic equilibrium.

Finally, we decompose the current \( J\subsc{ext} = \psi^+ - \psi^-\) into the apparent fluxes
\nw{
\begin{equation}
\begin{aligned}
  \psi^+ &= \frac{L_{\ce{E}} \xi(\vec{y})}{N_{\ce{E}}(\vec{y})} k_{1}k_{4}k_{5}k_{6} [\ce{S1}] [\ce{S2}]
  > 0\,,
  \\
 \psi^- &=\frac{L_{\ce{E}} \xi(\vec{y})}{N_{\ce{E}}(\vec{y})} k_{-1}k_{-4}k_{-5}k_{-6} [\ce{P}]
 > 0\,.
\end{aligned}
\label{eq:apparent-fluxes--single-cycle}
\end{equation}
Here, it is important to note that while
\begin{align*}
  \psi^{+} - \psi^{-} = k_6 [\ce{EP}] - k_{-6} [\ce{E}] [\ce{P}]\,,
\end{align*}
there are several cancellations happening in the derivation of \Eq{expl--open-steady-current} implying that
\begin{align*}
  \psi^{+} &\neq k_6 [\ce{EP}]\,, & \psi^{-} \neq k_{-6} [\ce{E}] [\ce{P}]\,.
\end{align*}
We elaborate on these cancellations in this special case in appendix~\ref{sec:appendix--ratelaws--enzyme} as well as for the general case in appendix~\ref{sec:appendix--fluxes--generic}.
}

\subsubsection*{Flux--force relation}

With the explicit expressions for the net force, \Eq{net-delta-G}, and the apparent fluxes, \Eq{apparent-fluxes--single-cycle}, of the emergent cycle we explicitly verify the flux--force relation at the coarse-grained level:
\begin{align*}
  RT \ln \frac{\psi^+}{\psi^-}
  = RT \ln \frac{k_{1}k_{4}k_{5}k_{6}[\ce{S1}][\ce{S2}]}{k_{-1}k_{-4}k_{-5}k_{-6}[\ce{P}]}
  = -\Delta\subsc{ext}G\,.
\end{align*}
This flux--force relation implies that the reaction current is always aligned with the net force along this reaction: \(J\subsc{ext} > 0 \Leftrightarrow -\Delta\subsc{ext}G > 0\).
In other words, the reaction current directly follows the force acting on this reaction.

In fact, in this case we can connect the flux--force relation to the second law of thermodynamics.
The EPR reads
\begin{align*}
  T \EP(\vec{x}\subsc{ss}, \vec{y}) &= - \vec{J}\subsc{ss} \cdot \Delta\subsc{r} \vec{G}
  = - J\subsc{int} \, 
    \vec{C}\subsc{int} \cdot\Delta\subsc{r} \vec{G}
    - J\subsc{ext} \, \vec{C} \subsc{ext} \cdot \Delta\subsc{r} \vec{G} \\
  &= - J\subsc{ext}\, \Delta\subsc{ext} G =  RT (\psi^+ - \psi^-) \ln \frac{\psi^+}{\psi^-}\geq 0\,.
\end{align*}
With this representation, it is evident that the flux--force relation ensures the second law: \(\EP \geq 0\).
Moreover, we see explicitly that the EPR is faithfully reproduced at the coarse-grained level.
This shows the thermodynamic consistency of our coarse-graining procedure.

\subsection{Active Membrane Transport}
\label{sec:transporter}

\begin{figure}[t]
  \centering
  \includegraphics{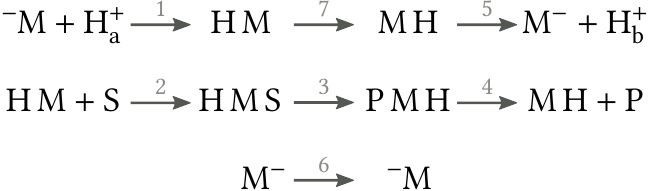}
  \caption{
  \new{Reaction graph for the mechanism} modeling the active transport of protons from one side of a membrane, \(\ce{H^+_a}\), to the other side, \(\ce{H^+_b}\).
    The transport is coupled to the catalysis of a substrate, \(\ce{S}\), to a product, \(\ce{P}\).
    The free transporter itself exists in two different conformations denoted \(\ce{^-M}\) and \(\ce{M-}\), respectively.
    Again, all reactions are considered reversible and to follow mass--action kinetics.
    A reference forward direction is indicated as arrows from left to right.
  }
  \label{fig:transporter--closed}
\end{figure}

\new{We now turn to the second example introduced in Fig.~\ref{fig:cg}: A membrane 
protein, \(\ce{M}\), that models a proton pump similar to the one presented in Ref.~\cite{Pietrobon.Caplan1985}.
It transports} protons from one side of the membrane (side a) to the other (side b).
The membrane protein itself is assumed to be charged to facilitate binding of the protons and to have different conformations \(\ce{M-}\) and \(\ce{^-M}\) where it exposes the binding site to the two different sides of the membrane.
Furthermore, when a proton is bound, it can either bind another substrate \(\ce{S}\) when exposing the proton to side a -- or the respective product \(\ce{P}\) when the proton is exposed to side b.
The latter could be some other ion concentrations on either side of the membrane -- or an energy rich compound (\(\ce{ATP}\)) and its energy poor counterpart (\(\ce{ADP}\)).
The reactions modelling this mechanism are summarized \new{in the reaction graph} in \Fig{transporter--closed}.

In order to find a coarse-grained description for this transporter we apply our result.
\new{Since the procedure is already detailed in example~\ref{sec:enzyme}, we omit some repetitive explanations in this example.}

\subsubsection*{(1) Closed system -- internal cycles}

This closed system has no cycle, therefore Wegscheider's conditions do not impose any relation between the reaction rate constants.
There are three conservation laws in the closed system,
\vspace{-1em}
\begin{align*}
  \vec{\ell}_{\ce{M}} &= \kbordermatrix{
    & \cr
    \color{g}\ce{^-M} 	&1 \cr
    \color{g}\ce{H M} 	&1 \cr
    \color{g}\ce{H M S}	&1 \cr
    \color{g}\ce{P M H}	&1 \cr
    \color{g}\ce{M H} 	&1 \cr
    \color{g}\ce{M-} 	&1 \cr
    \color{g}\ce{H^+_a}  &0 \cr
    \color{g}\ce{H^+_b}  &0 \cr
    \color{g}\ce{S}  	&0 \cr
    \color{g}\ce{P}  	&0 \cr
  }\,,
  &
  \vec{\ell}_{H} &= \kbordermatrix{
    & \cr
    \color{g}\ce{^-M} 	&0 \cr
    \color{g}\ce{H M} 	&1 \cr
    \color{g}\ce{H M S}	&1 \cr
    \color{g}\ce{P M H}	&1 \cr
    \color{g}\ce{M H} 	&1 \cr
    \color{g}\ce{M-} 	&0 \cr
    \color{g}\ce{H^+_a}  &1 \cr
    \color{g}\ce{H^+_b}  &1 \cr
    \color{g}\ce{S}  	&0 \cr
    \color{g}\ce{P}  	&0 \cr
  }\,,
  &
  \vec{\ell}_{S} &= \kbordermatrix{
    & \cr
    \color{g}\ce{^-M} 	&0 \cr
    \color{g}\ce{H M} 	&0 \cr
    \color{g}\ce{H M S}	&1 \cr
    \color{g}\ce{P M H}	&1 \cr
    \color{g}\ce{M H} 	&0 \cr
    \color{g}\ce{M-} 	&0 \cr
    \color{g}\ce{H^+_a}  &0 \cr
    \color{g}\ce{H^+_b}  &0 \cr
    \color{g}\ce{S}  	&1 \cr
    \color{g}\ce{P}  	&1 \cr
  }\,.
\end{align*}
They represent the conservation of membrane protein \new{(\(L_{\ce{M}}\))}, proton \new{(\(L_{\ce{H}}\))}, and substrate moieties \new{(\(L_{\ce{S}}\))}, respectively, showing that these three are conserved independently.
For any initial condition, the corresponding rate equations will relax to a unique steady-state solution satisfying thermodynamic equilibrium, \(\vec{J}(\vec{z})=0\).

\subsubsection*{(2) Open system -- emergent cycles}

\begin{figure}[t]
  \centering
  \includegraphics[scale=1]{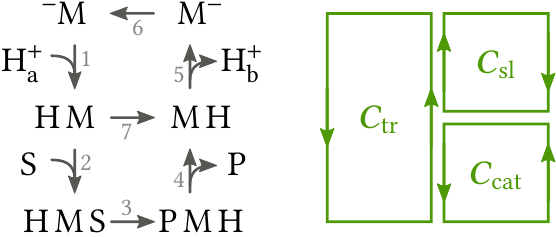}
  \caption{
    [left] Active membrane transport as \new{a graph representing the \emph{open}} chemical network.
    The proton concentrations \(\ce{H^+_a}\) and \(\ce{H^+_b}\), as well as the substrate and the product are chemostatted, thus are associated to the edges of the graph.
    [right] Graphical representations for the three \new{distinct cycles in this graph}.
    Only two of them are independent and we choose \(\vec{C}\subsc{\nw{cat}}\) and \(\vec{C}\subsc{sl}\) as a basis in the main text.
    The third is their \nw{difference \(\vec{C}\subsc{tr} = \vec{C}\subsc{cat} - \vec{C}\subsc{sl}\).}
  }
  \label{fig:transporter--open}
\end{figure}

We now fix the concentrations of the protons \(\ce{H^+_a}\) and \(\ce{H^+_b}\) in the two reservoirs, as well as the substrate and the product concentrations.
The reaction network for this open system is depicted in \Fig{transporter--open}.
The open system still has a conserved membrane protein moiety while the conservation laws of protons and substrate are broken upon chemostatting.
Furthermore, there are two emergent cycles now,
\vspace{-1em}
\begin{align}
  \nw{
    \vec{C}\subsc{cat}} &\nw{= 
  \kbordermatrix{
    & \cr
    \color{g}1& 0\cr
    \color{g}2& 1\cr  
    \color{g}3& 1\cr  
    \color{g}4& 1\cr    
    \color{g}5& 0\cr   
    \color{g}6& 0\cr
    \color{g}7& -1      
  }} & &\text{and} &
  \vec{C}\subsc{sl} &=
  \kbordermatrix{
    & \cr
    \color{g}1& -1\cr
    \color{g}2& 0\cr  
    \color{g}3& 0\cr  
    \color{g}4& 0\cr    
    \color{g}5& -1\cr   
    \color{g}6& -1\cr
    \color{g}7& -1      
  }\,.
\end{align}
Their visual representation \new{as circuits} is given on the right of \Fig{transporter--open}.

\subsubsection*{(3) Net Stoichiometry and net forces}

The first emergent cycle has the net stoichiometry
\nw{
\( \ce{S} \rightleftharpoons \ce{P} \), which represents pure catalysis with net force 
\begin{align}
  -\Delta\subsc{cat}G = RT \ln \frac{k_{2}k_{3}k_{4}k_{-7}[\ce{S}]}{k_{-2}k_{-3}k_{-4}k_{7}[\ce{P}]} \,.
  \label{eq:transporter--net-force--enzymatic}
\end{align}
}
The second cycle has net stoichiometry \(\ce{H^+_b}\rightleftharpoons\ce{H^+_a}\).
This represents the slip of one proton from side \(b\) back to side \(a\) with net force
\begin{align}
  -\Delta\subsc{sl}G &= RT \ln \frac{k_{-1}k_{-5}k_{-6}k_{-7}[\ce{H^+_b}]}{k_{1}k_{5}k_{6}k_{7}[\ce{H^+_a}]}\,.
  \label{eq:transporter--net-force--slip}
\end{align}
For later reference, we note that the
\nw{
  difference \(\vec{C}\subsc{tr}= \vec{C}\subsc{cat} - \vec{C}\subsc{sl}\) has net stoichiometry
\(\ce{H^+_a + S} \rightleftharpoons \ce{H^+_b + P}\).
This is the active transport of a proton from side \(a\) to side \(b\), under catalysis of one substrate into one product.
The net force of this reaction is
\begin{align}
  -\Delta\subsc{tr}G &= -\Delta\subsc{cat}G + \Delta\subsc{sl}G \nonumber\\ &= RT \ln \frac{k_{1}k_{2}k_{3}k_{4}k_{5}k_{6}[\ce{H^+_a}][\ce{S}]}{k_{-1}k_{-2}k_{-3}k_{-4}k_{-5}k_{-6}[\ce{H^+_b}][\ce{P}]} \,.
  \label{eq:transporter--net-force--transport}
\end{align}
}

\subsubsection*{(4) Apparent Fluxes}

\nw{Solving the linear rate equations (see appendix~\ref{sec:appendix--steady-state}),}
we have a solution for the steady-state concentrations.
The exact expressions are given in appendix~\ref{sec:appendix--steady-state--transporter}.
With the steady-state concentrations, we calculate the contributions of both cycles to the steady-state current:
  \(
  \vec{J}\left( \vec{x}\subsc{ss}, \vec{y} \right) = J\subsc{cat}\,\vec{C}\subsc{cat} + J\subsc{sl}\,\vec{C}\subsc{sl} 
  \).
  \nw{Each current contribution is given by a single reaction:}
  \nw{
\begin{align*}
  J\subsc{cat} &= J_{2} =: \psi^{+}\subsc{cat} - \psi^{-}\subsc{cat}\,, & 
  J\subsc{sl} &= -J_{1}
  =: \psi^{+}\subsc{sl} - \psi^{-}\subsc{sl} \,.
\end{align*}
}
With the abbreviations
\begin{align*}
  \xi\subsc{cat} &:= k_{-6}k_{-5}[\ce{H^+_b}] + k_{1}k_{-5}[\ce{H^+_a}][\ce{H^+_b}] + k_{6}k_{1}[\ce{H^+_a}]\,,\\
  \xi\subsc{sl} &:=  k_{3}k_{4} + k_{-2}k_{4} + k_{-3}k_{-2}\,, 
\end{align*}
we can express the apparent fluxes as
\nw{
\begin{align*}
  \frac{N\subsc{M}}{L_{\ce{M}}} \psi^+\subsc{cat}
  &=
  k_{1}k_{2}k_{3}k_{4}k_{5}k_{6}[\ce{H^+_a}][\ce{S}]
   + \xi\subsc{cat} k_{-7}k_{2}k_{3}k_{4}[\ce{S}]
  \,, \\
  \frac{N\subsc{M}}{L_{\ce{M}}} \psi^{-}\subsc{cat}
  &=
  k_{-1}k_{-2}k_{-3}k_{-4}k_{-5}k_{-6}[\ce{H^+_b}][\ce{P}]
  + \xi\subsc{cat} k_{7}k_{-2}k_{-3}k_{-4}[\ce{P}]
  \,,\\
  \frac{N\subsc{M}}{L_{\ce{M}}} \psi^{+}\subsc{sl}
  &=
  k_{-1}k_{-2}k_{-3}k_{-4}k_{-5}k_{-6}[\ce{H^+_b}][\ce{P}]
  +
  \xi\subsc{sl} k_{-1}k_{-5}k_{-6}k_{-7}[\ce{H^+_b}]
  \,, \\ 
  \frac{N\subsc{M}}{L_{\ce{M}}} \psi^{-}\subsc{sl}
  &=
  k_{1}k_{2}k_{3}k_{4}k_{5}k_{6}[\ce{H^+_a}][\ce{S}]
  +
  \xi\subsc{sl} k_{1}k_{5}k_{6}k_{7}[\ce{H^+_a}]
  \,,
\end{align*}
The derivation for these equations is detailed in appendix~\ref{sec:appendix--ratelaws--transporter}.
Note that \(N_{\ce{M}}\) depends on all rate constants and all chemostat concentrations.
}

\subsubsection*{Breakdown of the flux--force relation}

We see that the abbreviated terms \(\xi\) appear symmetrically in the forward and backward fluxes.
Therefore, when the net forces are zero, necessarily the currents vanish and the system is at thermodynamic equilibrium.
However, in general, the currents do not vanish.
Moreover, the concentrations of the chemostats appear in the four different fluxes in different combinations -- indicating that both net forces couple to both coarse-grained reactions.
Due to this coupling, it is impossible to find nice flux--force relations for the two reactions independently:
\new{
  \begin{align}
    \nw{-\Delta\subsc{cat}G} &\nw{\neq RT \ln \frac{\psi^{+}\subsc{cat}}{\psi^{-}\subsc{cat}}\,,} & 
    -\Delta\subsc{sl}G &\neq RT \ln \frac{\psi^{+}\subsc{sl}}{\psi^{-}\subsc{sl}}\,. 
  \end{align}
}
To the contrary, it is \nw{easy} to find concentrations for the four chemostats where the \nw{catalytic} force is so strong that it drives the \nw{slip} current againts its natural direction -- giving rise to a negative contribution in the entropy-production rate.
Nonetheless, the overall EPR is correctly reproduced at the coarse-grained level:
\begin{align*}
  T \EP &= \new{-\vec{J}\subsc{ss} \cdot \Delta\subsc{r} \vec{G}} = - J\subsc{cat} \,\Delta\subsc{cat}G - J\subsc{sl}\,\Delta\subsc{sl}G \geq 0\,.
\end{align*}
Since this is\new{, by construction,} the correct entropy-production rate of the full system at steady state, we know that it is always non-negative -- and that the coarse-graining procedure is thermodynamically consistent.
This example shows explicitly that biochemical reaction networks need not satisfy the flux--force relation, nor need their currents and forces be aligned to comply with the second law.
\nw{After all, the function of this membrane protein is to transport protons from side a to side b against the natural concentration gradient.}








\section{Cycle-based Coarse Graining}

From the perspective of a single biocatalyst, the rest of the cell (or cellular compartment) serves as its environment, providing a reservoir for different chemical species.
Our coarse graining exploits this perspective to disentangle the interaction of the catalyst with its environment -- in the form of emergent cycles -- from the behavior of the catalyst in a (hypothetical) closed box at thermodynamic equilibrium -- in the form of the internal cycles.
From the perspective of the environment, only the interactions with the catalyst matter, i.e.\@ the particle exchange currents: \new{They prescribe the substrate/product turnover and when combined with the reservoir's concentrations (chemical potentials) also the dissipation.
Our coarse graining respects the reservoir's concentrations and incorporates all the emergent cycles that exchange particles with the reservoir.
It thus correctly reproduces the exchange currents:} This is the fundamental reason why we can replace the actual detailed mechanism of the catalyst with a set of coarse-grained reactions \new{in a thermodynamically exact way}.
A formal version of this reasoning, including all necessary rigor and a constructive prescription to find the apparent fluxes, is provided in \new{appendix~\ref{sec:appendix--fluxes}}.

In our examples we illustrated the fundamental difference between the case where a catalyst can be replaced with a single coarse-grained reaction and the case where this is not possible.
In the first case, such a catalyst interacts with substrate and product molecules that are \new{coupled via exchange of mass in a specific stoichiometric ratio.
This is known as tight coupling.}
Whether or not the catalysis is additionally modified by activators or inhibitors, does not interfere with this condition.
After all, the modifiers are neither consumed nor produced.
Thus they appear only in the normalizing denominators of the steady-state concentrations and affect the kinetics while leaving the thermodynamics untouched.
Furthermore, if there is only one single emergent cycle in a catalytic mechanism, any product of pseudo-first-order rate constants along any \new{circuit} in the network will either (i) satisfy Wegscheider's conditions or (ii) reproduce (up to sign) the net force, \(-\Delta_{\alpha}G\), of the emergent cycle.
Ultimately, this is why the flux--force relation holds in this \new{tightly-coupled} case.
A formal version of this proof, including all necessary rigor, is provided in \new{appendix~\ref{sec:appendix--flux-force}.} 

In the case where we have to provide two or more coarse-grained reactions, the catalytic mechanism couples several processes that are not \new{tightly coupled via exchange of mass.
To the contrary: The turnover of different substrates/products need not have fixed stoichiometric ratios.
In fact, their ratios will depend on the environment's concentrations.}
In this case the flux--force relation does not hold in general, as we proved with our counter-example.
After all, when several processes are coupled, the force of one process can overcome the force of the second process to drive the second current against its natural direction.
This transduction of energy~\new{\cite{Hill1977a, Altaner.etal2015}} would not be possible at a coarse-grained level, if the flux--force relation was always true.

\new{
  We now asses the reduction provided by our procedure:
The number \(C\) of coarse-grained reactions \(\alpha\) is always lower than the number \(M\) of reaction steps \(\rho\) in the original mechanism.
This can be understood from the graph representation of the open system:
The number \(B\) of circuits in a connected graph is related to its number \(N\)  of vertices (catalyst states) and the number \(M\) of edges (reaction steps) by \(B = M - N + 1\)~\cite{Tutte2001}.
Some of the circuits represent internal cycles, rendering \(B\) an upper bound to the number of emergent cycles \(C\).
Since the number \(N\) of catalyst states is at least two, these numbers are ordered: \(M > B \geq C\).
This proves that our coarse graining always reduces the number of reactions.
}

\section{Discussion}

\new{
  The original work of Michaelis and Menten~\cite{Michaelis.Menten1913} was based on a specific enzyme that converts a single substrate into a single product assuming a totally irreversible step.
  Their goal was to determine the rate of production of product molecule.
  Later progress in enzyme kinetics extended their method to deal with fully reversible mechanisms, as well as many substrates, many products and modifiers~\cite{Cornish-Bowden2013}.
  The focus on the turnover led many people to identify the net effect of the enzyme with a \emph{single} effective reaction, describing its kinetics with the Michaelis--Menten equation (or one of its generalizations). 
  Our coarse-graining indeed incorporates all these special cases:
  The Michaelis--Menten equation arises from coarse graining a mechanism of the form
  \begin{align}
    \ce{S + E} \rightleftharpoons \ce{ES} \rightleftharpoons \ce{EP} \rightleftharpoons \ce{E + P}
    \label{eq:michaelis--menten}
  \end{align}
  and assuming that the last reaction step, the release of the product, is much faster than the other steps.
  Then the coarse-grained reaction current is identical to the substrate/product turnover.
  Importantly, our procedure highlights that there is no direct correspondence between the number of required net reactions and the number of circuits in the reaction graph -- even of the open system.
  Some circuits correspond to \emph{internal cycles} that play a kinetic role, not leaving a trace in the thermodynamic forces.
  Only the \emph{emergent cycles} need to be taken into account for the coarse graining.
  Thus the net effect of a multi-cyclic catalyst might be consistently expressed as a single effective reaction, as seen in the example~\ref{sec:enzyme}.

  \nw{Likewise,} in theoretical studies of biochemical systems, effective unimolecular reactions of the form
  \begin{align*}
    \ce{A} \stackrel{\ce{X}\qquad\ce{Y}}{\vcenter{\hbox{\includegraphics{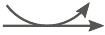}}}} \ce{B}
  \end{align*}
  are frequently used, where the reaction rate constants satisfy
  \begin{align*}
    \frac{k^{+}}{k^{-}} = \exp\left[ \frac{\mu^{\circ}_{\ce{A}} - \mu^{\circ}_{\ce{B}} + \mu_{\ce{X}} - \mu_{\ce{Y}}}{RT} \right]\,.
  \end{align*}
  Here, the chemical potentials, \(\mu\), account for the thermodynamic force exerted by \(\ce{X}\) and \(\ce{Y}\).
  Even when the actual effective reaction does not follow mass--action kinetics, this equation is assumed, implying that the effective reaction fluxes are \(k^{+}[A] = \psi^{+}\) and \(k^{-}[B] = \psi^{-}\), and the ``constants'' \(k\) indeed depend on some concentrations.
  This is only consistent if the implicit conversion mechanism is tightly coupled by exchange of mass:
  When tightly coupled, the differences of the chemical potentials represent the Gibbs free energy change along the reaction \(\ce{A + X} \rightleftharpoons \ce{B + Y}\).
  In this case, the above equation is the flux--force relation.
  Otherwise, our coarse-graining procedure reveals that this is thermodynamically inconsistent:  If the implicitly modelled catalysis is not tightly coupled via the exchange of mass, there is a hidden thermodynamic driving force that is independent of the concentrations of \(\ce{A}\) and \(\ce{B}\), while the turnover of \(\ce{X}\)/\(\ce{Y}\) is not in a stoichiometric ratio to the turnover of \(\ce{A}\)/\(\ce{B}\).
  We have seen in example~\ref{sec:transporter} that the flux--force relation indeed does not hold in this case.

The failure of the flux--force relation in the non-tightly coupled case does not imply inconsistent thermodynamics.
Our coarse-graining procedure indeed deals with this case very easily.
The resulting fluxes and forces reproduce the entropy-production rate while sacrificing the flux--force relation.
The key difference to the original ideas in enzyme kinetics is that the substrate/product turnover is split into \emph{several} effective reactions with their own reaction fluxes and forces, reproducing the entropy-production rate.
\nw{This is especially important for complex catalysts: Many models for molecular motors and active transporters are not tightly coupled.
These free-energy transducers often display slippage via futile cycles.
While some enzymes also show signs of slippage, many simple enzymes are modelled as tightly coupled -- which implies they satisfy the flux--force relation.
Our} coarse graining \nw{deals with all these cases and in that sense} goes far beyond Michaelis--Menten.

Our procedure greatly reduces the number of species and reactions involved in a network while reproducing the entropy-production rate.
  This comes at the cost of complicated effective fluxes (rate laws).
  They are rational functions of the involved concentrations and thus more complicated than simple mass--action kinetics.
  Nonetheless, our procedure is constructive by giving these complicated expressions explicitly.
  With the explicit solutions at hand, further assumptions can be made to simplify the effective fluxes -- as in the case of the original Michaelis--Menten equation.
  Note that these additional simplifications may have an impact on the entropy-production rate, in the worst case breaking the thermodynamic consistency.
  This trade-off between simplicity and thermodynamic correctness needs to be evaluated case by case.

We now discuss the limitations of our approach.
The presented coarse-graining procedure is exact in \emph{steady-state} situations, arbitrarily far from equilibrium.
When the surrounding reaction network is not in a steady state, the coarse graining can still be used:  Then the coarse-grained reaction fluxes and forces have to be considered \emph{instantaneous} -- \nw{they} change in time due to the changing substrate/product (or modifier) concentrations.
  Underlying this point of view is a separation of time scales:
  When the abundance of substrates and products is very large, as compared to the abundance of catalyst, then the concentrations of the latter change much more quickly.
  This results in a quasi-steady state for the catalyst-containing species.
Consequently, our coarse graining cannot capture the contribution to dissipation that arises in this fast relaxation dynamics.
It only captures the dissipation due to the conversion of substrate into product.
This reasoning can be made more rigorous: There are time-scale separation techniques for deterministic rate equations~\cite{Segel.Slemrod1989, Lee.Othmer2009} frequently used in biochemical contexts~\cite{Gunawardena2014}, furthermore stochastic corrections due to small copy-numbers~\cite{Thomas.etal2012} and even effective memory effects~\cite{Rubin.etal2014, Rubin.Sollich2016} can be incorporated.
However, these techniques do not explicitly address the question of thermodynamic consistency and we think that combining our coarse-graining with these techniques is a promising endeavour for the future.

We restricted the entire reasoning in this paper to catalysts.
They follow linear rate equations when their reaction partners have constant concentrations.
This linearity allowed us to give explicit solutions for general catalysts.
Focusing on the emergent cycles to reproduce the correct thermodynamics paves the way to apply a similar procedure beyond catalysts:
Reaction networks that remain non-linear after chemostatting still have emergent cycles~\cite{Polettini.Esposito2014}. \nw{They can be calculated algebraically from bases for the nullspaces of the full and the reduced stoichiometric matrices, \(\stoich\) and \(\stoich^{X}\).
The cycles in non-linear networks may not have a representation as circuits in the reaction graph, as we have seen with the internal cycle of the enzyme in a closed box.
Nonetheless,} each of the emergent cycles \nw{\(\vec{C}_{\alpha}\)} can serve as an effective reaction\nw{: It has a well defined stoichiometry, \(\stoich^{Y}\vec{C}_{\alpha}\), and a well defined net force, \(-\Delta\subsc{r}\vec{G}\cdot \vec{C}_{\alpha}\)}.
The steady state \nw{concentrations as well as the fluxes, however,} need to be determined case by case.
Non-linear differential equations can be multi-stable, where our coarse graining applies to each stable steady state.
Some non-linear ODEs exhibit limit cycles, thus never reaching a steady state.
In this case our procedure is no longer applicable.
}

\section{Summary}

We have presented a coarse-graining procedure for biocatalysts and have shown that it is thermodynamically consistent.
During this coarse graining procedure, a detailed catalytic mechanism is replaced by a few net reactions.
The stoichiometry, \new{deterministic kinetic rate laws} and net forces for the coarse-grained reactions are calculated \new{explicitly} from the detailed mechanism -- ensuring that at steady state the detailed mechanism and the net reactions have \new{both the same substrate/product turnover and} the same entropy-production rate.

Furthermore, we have shown that in the \new{tightly-coupled} case where a detailed mechanism is replaced by a single reaction, this net reaction satisfies a flux--force relation.
In the case where a detailed mechanism has to be replaced with several net reactions, the flux--force relation does not hold for the net reactions due to cross-coupling \new{of independent thermodynamic forces}.
Ultimately, this cross-coupling allows the currents and forces not to be aligned -- while complying with the second law of thermodynamics.

Overall, we have shown that coarse-graining schemes which preserve the correct thermodynamics far from equilibrium are not out of reach.

\begin{acknowledgments}
  This work is financially supported by the National Research Fund of Luxembourg in the frame of AFR PhD Grants No. 7865466 and No. 9114110.
  Furthermore, this research is funded by the European Research Council project NanoThermo (ERC-2015-CoG Agreement No. 681456).

\end{acknowledgments}


\subsubsection*{Conflict of Interest}
The authors declare that they have no conflict of interest.

\newpage
\appendix

\section{Diagrammatic method for explicit steady states of linear reaction networks}
\label{sec:appendix--steady-state}


We consider a catalytic mechanism with a catalyst and several substrates, products, inhibitors or activators.
The mechanism is resolved down to elementary reactions following mass--action kinetics.

Upon chemostatting all the substrates, products, inhibitors and activators -- summarized as \(\vec{y}\) --  we are left with rate equations that are linear in the catalyst-containing species -- summarized as \(\vec{x}\).
\new{
While the steady-state equations alone, \(\vec{0} = \stoich^{X}\vec{J}(\vec{x}, \vec{y})\), are under-determined and linearly dependent, the open system still has a conservation law for the total catalyst-moiety concentration \(L = \sum_i x_i\), which again is a linear equation.
We can replace the first line of the steady-state equations with this constraint to arrive at linear equations \(L\vec{e}_1 = \mathbb{M}(\vec{y}) \vec{x}\), where \(\vec{e}_1 = \left( 1, 0, \dots \right)\) is the first Cartesian unit vector and \(\mathbb{M}(\vec{y})\) is an invertible square matrix that depends on the chemostat concentrations.
According to Cramer's rule the unique solution to this problem is given by
\begin{align}
  \frac{x^{*}_i}{L} = \frac{\det \mathbb{M}_i(\vec{y})}{\det\mathbb{M}(\vec{y})}\,,
  \label{eq:cramer-solution}
\end{align}
where \(\mathbb{M}_i(\vec{y})\) is identical to \(\mathbb{M}(\vec{y})\) just with the \(i\)th column replaced by \(\vec{e}_1\).
We now provide a diagrammatic method to represent this solution.
This diagrammatic method is frequently attributed to King and Altman~\cite{King.Altman1956} or Hill~\cite{Hill1966}, while an equivalent approach was already employed by Kirchhoff~\cite{Kirchhoff1847} to solve problems in electric networks.
We give the diagrammatic method in the language of graph theory~\cite{Tutte2001,Knauer2011}, for which we need some definitions.

The open pseudo-first-order reaction network has a simple representation as a connected graph \(\mathcal{G}\) where all the catalyst-containing species \(i\) form the vertices \(\mathcal{V}\) and the reactions \(\rho \cup -\rho\) form bidirectional edges \(\mathcal{R}\).
The reduced stoichiometric matrix \(\stoich^{X}\) is the \emph{incidence matrix} for this graph.

A closed self-avoiding path in a graph is a \emph{circuit} and can be identified with a vector \(c\in \mathbb{R}^{\mathcal{R}}\) over the edges, whose entries are in fact restricted to \(\{-1, 0, 1\}\).
Since a circuit is a closed path, it satisfies \(\stoich^{X}c=0\) and reaches as many vertices as it contains edges.
A graph not containing any circuit is called \emph{forest}, a connected forest is called \emph{tree}.

A connected subgraph \(\tau\subset\mathcal{G}\) is called \emph{spanning tree} if it spans all the vertices but contains no circuit.
The set \(\mathcal{T}\) of spanning trees of a finite graph is always finite.
A \emph{rooted spanning tree} is a tree where all the edges are oriented along the tree towards one and the same vertex, called the \emph{root}.

With these notions set, the determinants in \Eq{cramer-solution} can be written as
\begin{align*}
  \det \mathbb{M}_i(\vec{y}) &= \sum_{\tau \in \mathcal{T}_i} \prod_{\rho \in \tau} \tilde{k}_{\rho}(\vec{y})\,, \\
  \det \mathbb{M}(\vec{y}) &= \sum_{i} \sum_{\tau \in \mathcal{T}_i} \prod_{\rho \in \tau} \tilde{k}_\rho(\vec{y}) \eqqcolon N(\vec{y})\,.
\end{align*}
Here, \(\mathcal{T}_i\) is the set of spanning trees rooted in vertex \(i\), and \(\tilde{k}_\rho(\vec{y})\) is the pseudo-first-order rate constant of reaction \(\rho\).
Overall, Kirchhoff's formula for the solution to the linear problem is
\begin{align}
  \frac{x^{\ast}_{i}}{L}  &= \frac{1}{N(\vec{y})} \sum_{\tau \in \mathcal{T}_i} \prod_{\rho \in \tau} \tilde{k}_{\rho}(\vec{y}) \,.
  \label{eq:kirchhoff-formula}
\end{align}
From this result it is easy to confirm that the solution exists and is always unique as long as the chemostat concentrations are finite and positive.
Furthermore, the steady-state concentrations are expressed as sums of products of positive quantities, thus themselves are always positive.

While this formula is very compact and abstract, it is not obviously convenient for practical calculations.
However, the rooted spanning trees appearing in this formula can be visually represented as diagrams, as we will see in the following examples.
These diagrams are intuitive enough to make practical calculations with this formula feasible.
}
\subsection{Steady-state concentrations for the enzymatic catalysis}
\label{sec:appendix--steady-state--enzyme}
\new{
The enzymatic catalysis example in the main text, when open, is represented by the graph in figure~\ref{fig:expl-open}.
This graph has five vertices and six edges.
It contains three distinct circuits and twelve different spanning trees.

A visual representation of Kirchhoff's formula \eq{kirchhoff-formula} for its steady-state concentrations is given by the following diagrams:
}
\begin{align*}
  \frac{[\ce{E}]}{L_{\ce{E}}} = \frac{1}{N_{\ce{E}}} \nw{\left[\diagram{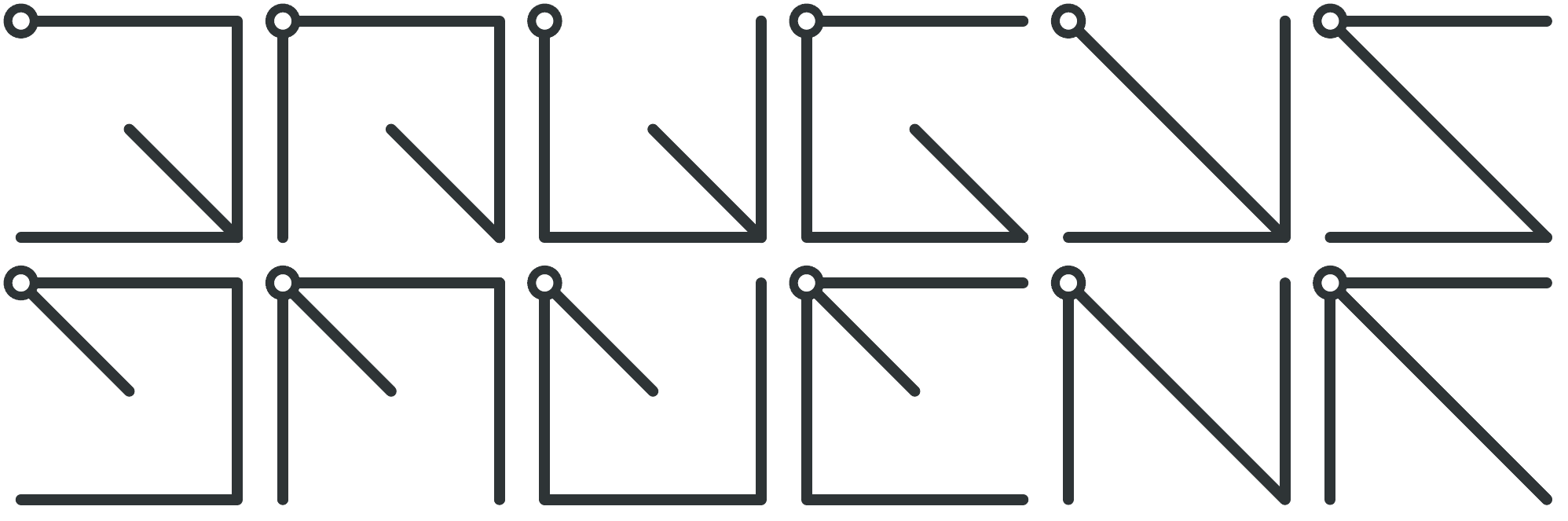}{0.2}\right]} \\
  \frac{[\ce{ES1}]}{L_{\ce{E}}} = \frac{1}{N_{\ce{E}}} \nw{\left[\diagram{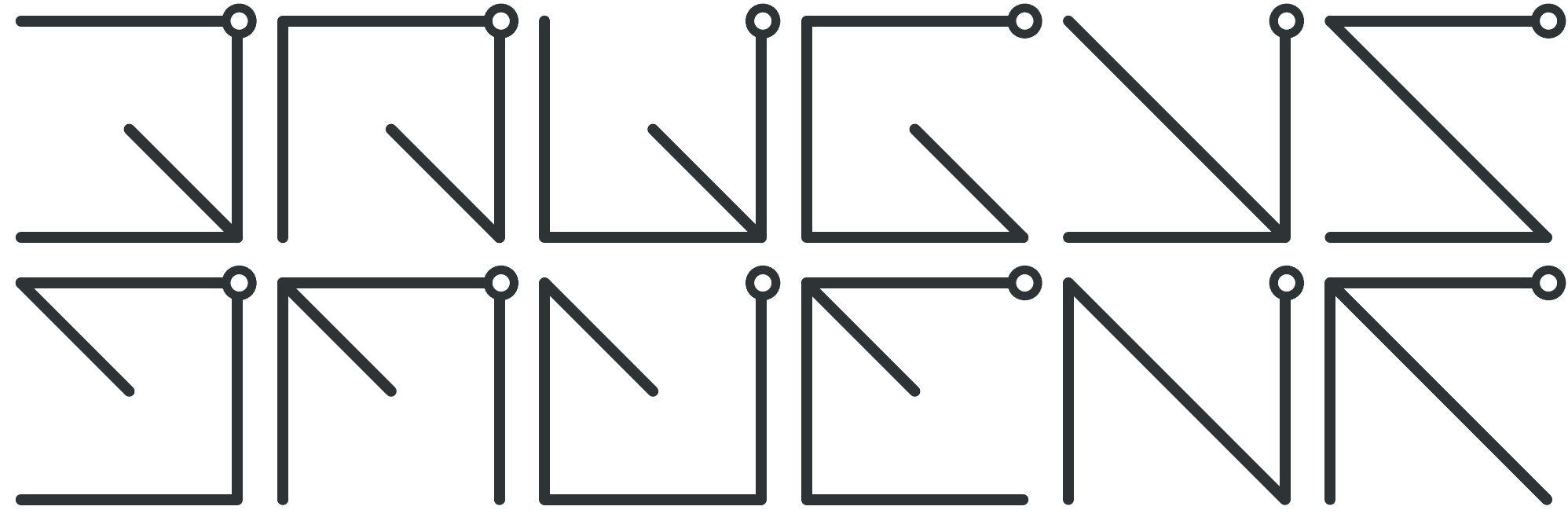}{0.2}\right]} \\
  \frac{[\ce{ES2}]}{L_{\ce{E}}} = \frac{1}{N_{\ce{E}}} \nw{\left[\diagram{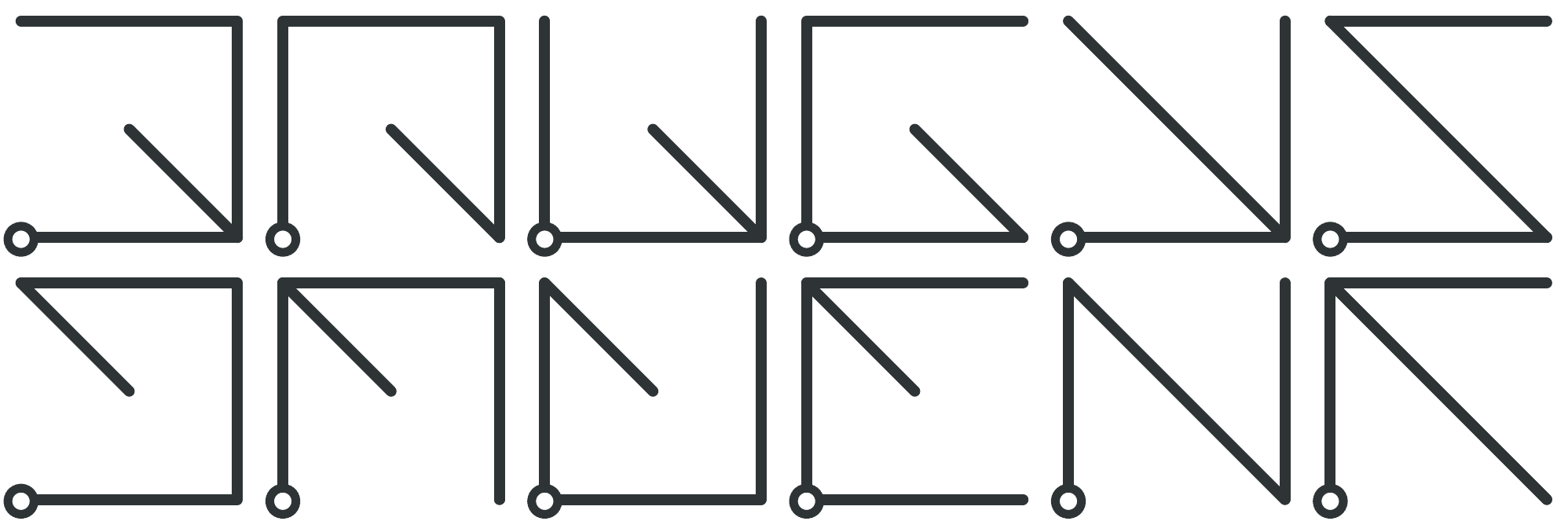}{0.2}\right]} \\
  \frac{[\ce{ES1S2}]}{L_{\ce{E}}} = \frac{1}{N_{\ce{E}}} \nw{\left[\diagram{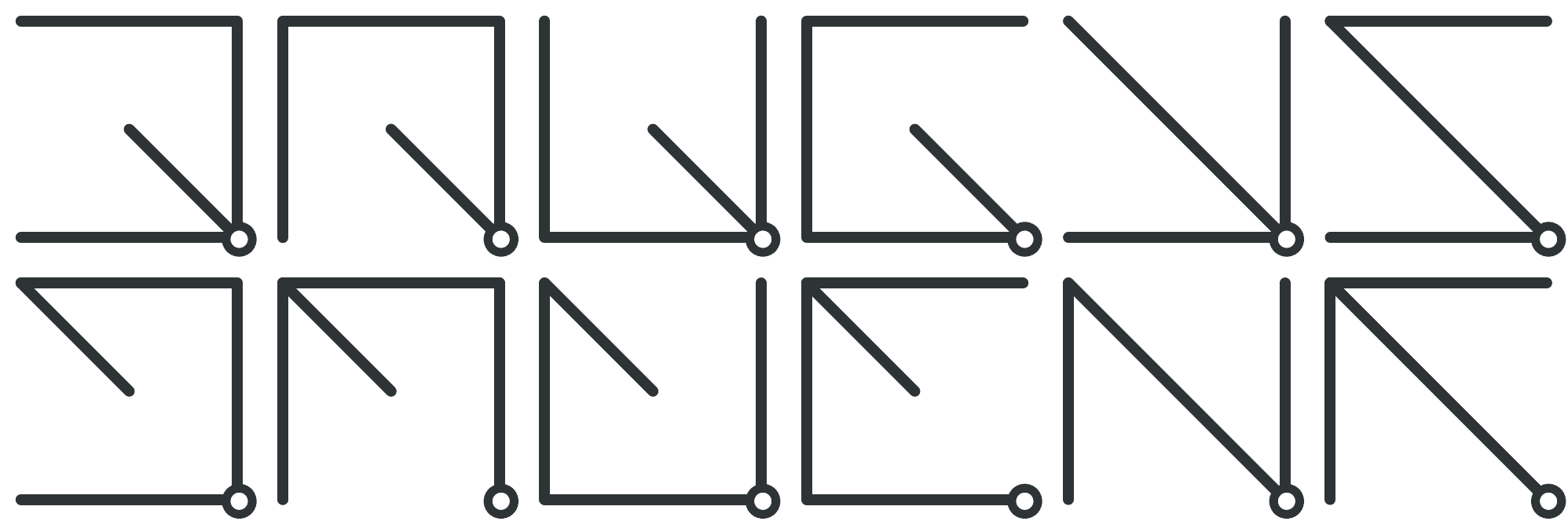}{0.2}\right]} \\
  \frac{[\ce{EP}]}{L_{\ce{E}}} = \frac{1}{N_{\ce{E}}} \nw{\left[\diagram{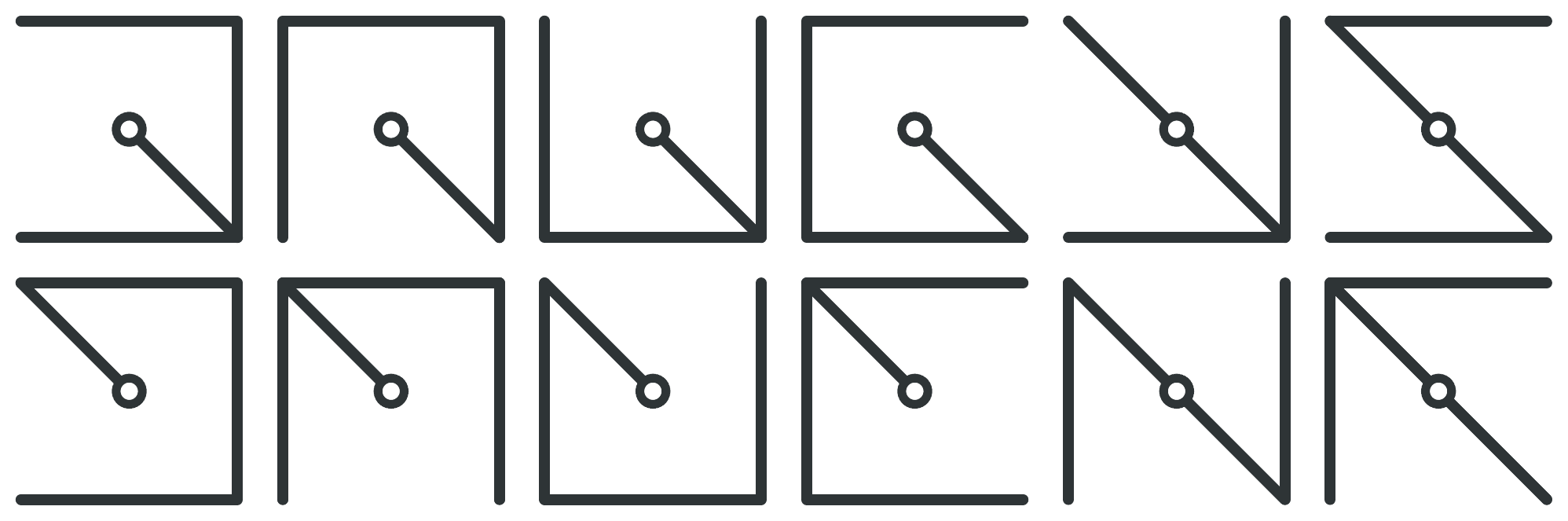}{0.2}\right]}
\end{align*}
\new{
Here, each diagram represents a product of pseudo-first-order rate constants over a spanning tree that is rooted in the (circled) vertex associated with the species we want to solve for (left hand side).
Thus, the concentrations are sums of twelve diagrams each, normalized by a denominator \(N_{\ce{E}}\) that equals the sum of all the 60 diagrams given above.
}
\subsection{Steady-state concentrations for the active transporter}
\label{sec:appendix--steady-state--transporter}

\new{
The active membrane transporter example in the main text, when open, is represented by the graph in figure~\ref{fig:transporter--open}.
This graph has six vertices and seven edges.
It contains three distinct circuits and 15 different spanning trees.

A visual representation of Kirchhoff's formula \eq{kirchhoff-formula} for its steady-state concentrations is given by the following diagrams:
}
\begin{align*}
  \frac{[\ce{^-M}]}{L_{\ce{M}}} = \frac{1}{N_{\ce{M}}} \nw{\left[\diagram{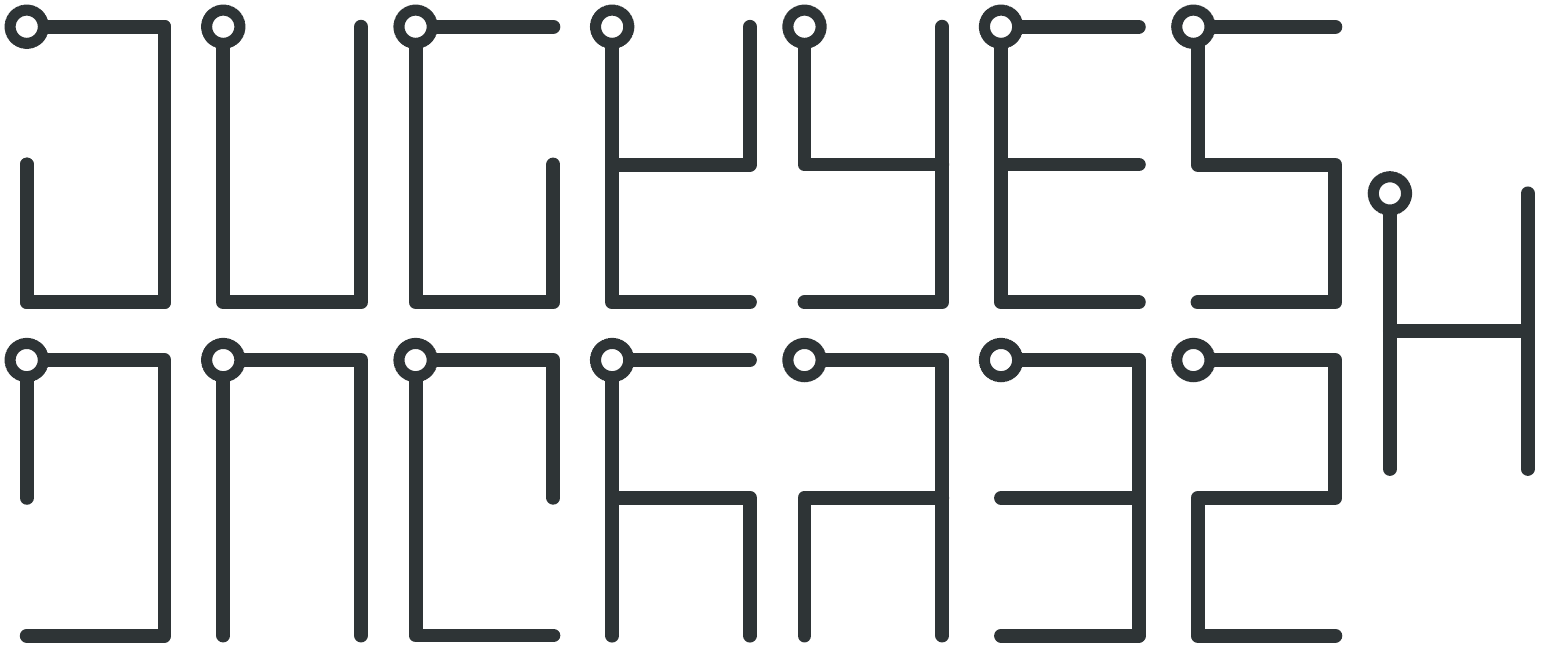}{0.2}\right]} \\
  \frac{[\ce{HM}]}{L_{\ce{M}}} = \frac{1}{N_{\ce{M}}} \nw{\left[\diagram{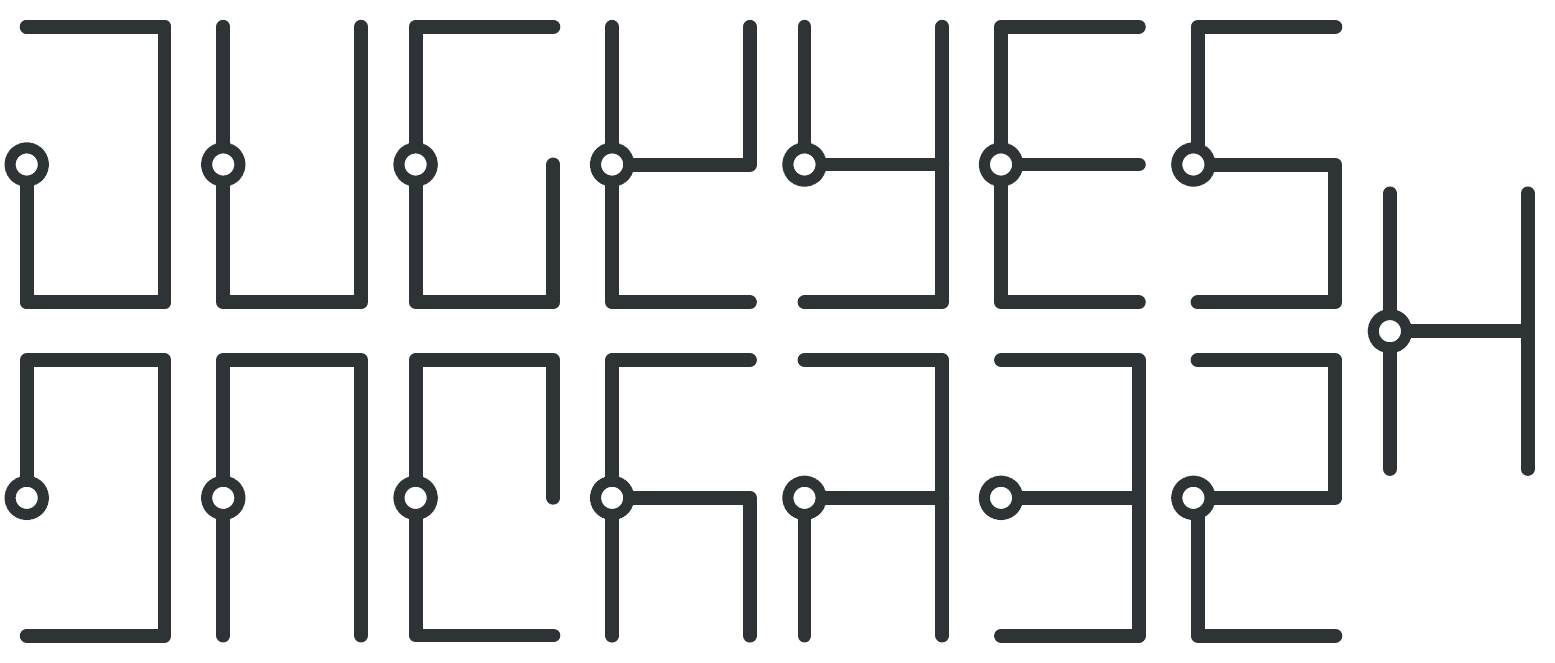}{0.2}\right]} \\
  \frac{[\ce{HMS}]}{L_{\ce{M}}} = \frac{1}{N_{\ce{M}}} \nw{\left[\diagram{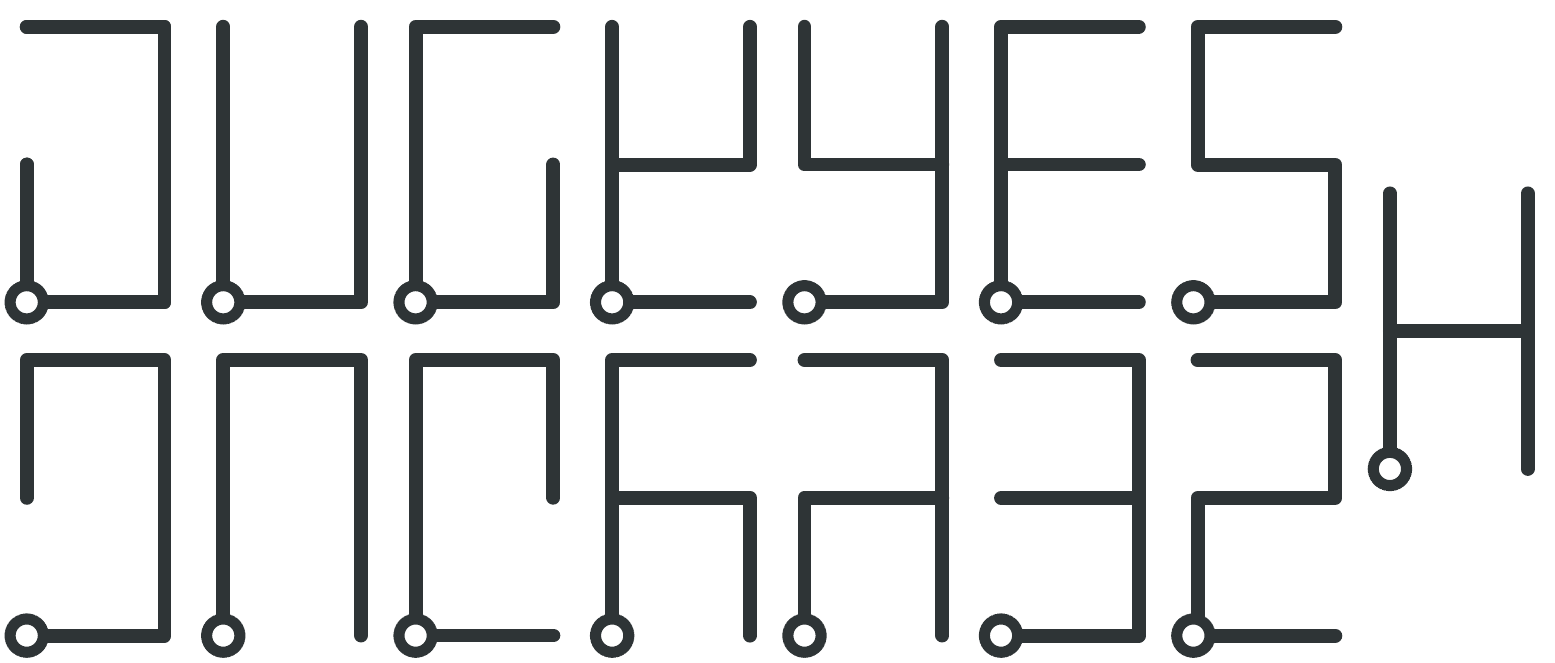}{0.2}\right]} \\
  \frac{[\ce{M-}]}{L_{\ce{M}}} = \frac{1}{N_{\ce{M}}} \nw{\left[\diagram{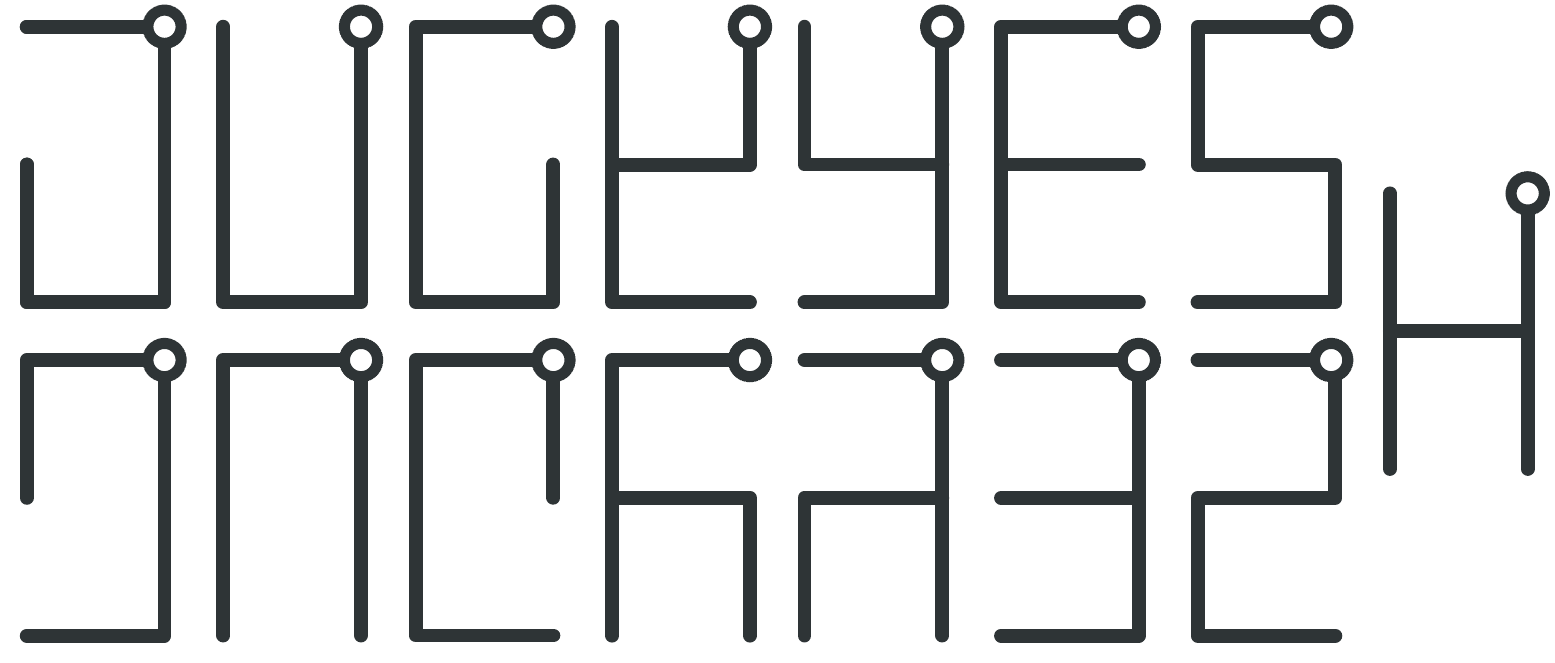}{0.2}\right]} \\
  \frac{[\ce{MH}]}{L_{\ce{M}}} = \frac{1}{N_{\ce{M}}} \nw{\left[\diagram{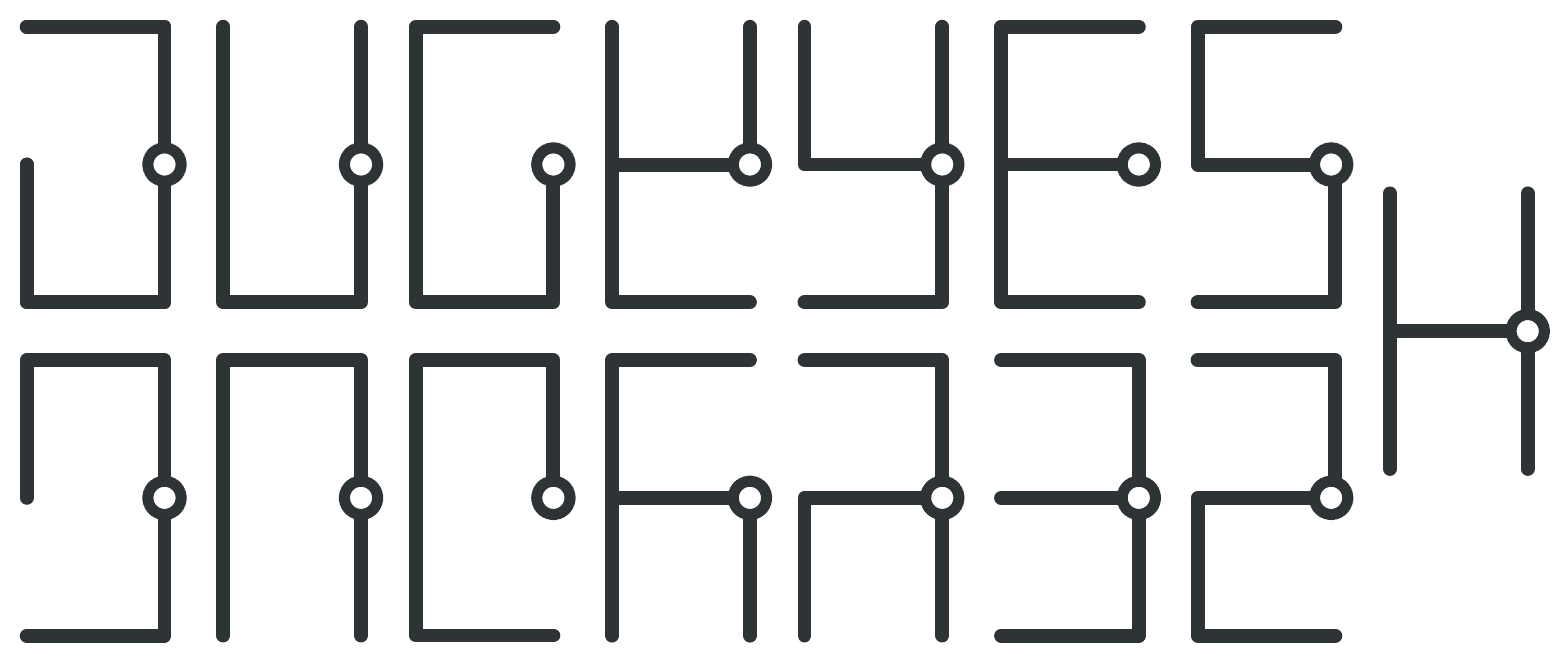}{0.2}\right]} \\
  \frac{[\ce{PMH}]}{L_{\ce{M}}} = \frac{1}{N_{\ce{M}}} \nw{\left[\diagram{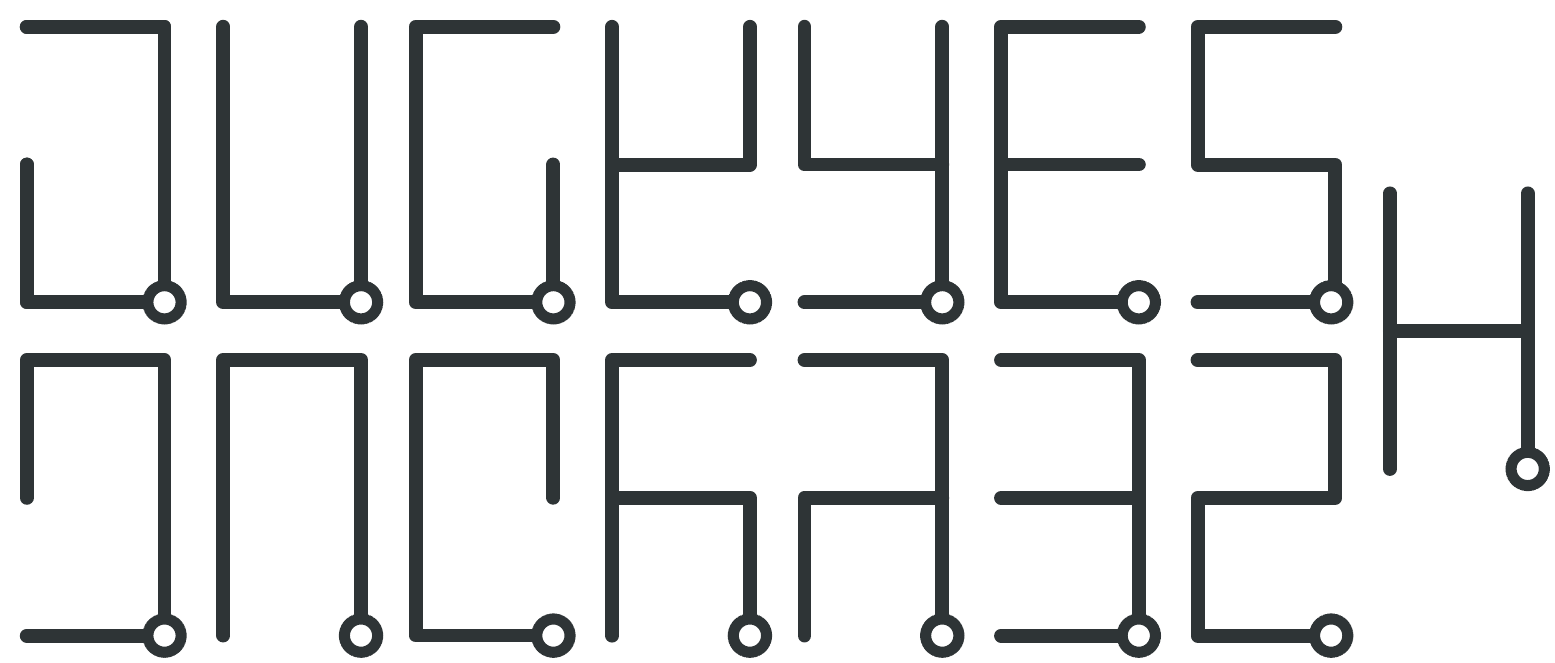}{0.2}\right]} \\
\end{align*}
\new{
Here, each diagram represents a product of pseudo-first-order rate constants over a spanning tree that is rooted in the (circled) vertex associated with the species we want to solve for (left hand side).
Thus, the concentrations are sums of 15 diagrams each, normalized by a denominator \(N_{\ce{M}}\) that equals the sum of all the 90 diagrams given above.
}

\section{Kinetic rate laws for the coarse-grained reactions}
\label{sec:appendix--fluxes}

\nw{
We now explicitly construct the kinetic rate laws as apparent cycle fluxes.
First, we make use of the diagrammatic method to derive the coarse-grained kinetic rate laws for the two example systems of the main text.
Then we generalize these examples to generic catalysts.
}

\nw{
\subsection{Kinetic rate laws for the enzymatic catalysis}
\label{sec:appendix--ratelaws--enzyme}
}

\nw{As shown in the main text, the cycle currents are
  \begin{align*}
    J\subsc{int} &= -J_{2} = k_{-2} [\ce{ES2}] - k_2 [\ce{E}][\ce{S2}]\,, \\
    J\subsc{ext} &= J_{6} = k_6 [\ce{EP}] - k_{-6} [\ce{E}] [\ce{P}]\,.
  \end{align*}
  Plugging in the diagrams (appendix \ref{sec:appendix--steady-state--enzyme}) for the steady-state concentrations of the enzyme-containing species we arrive at
  \begin{alignat*}{3}
    \frac{N_{\ce{E}}}{L_{\ce{E}}} J\subsc{int} &= & \qquad k_{-2} &\left[\diagram{enzyme--trees-ES2.pdf}{0.2}\right]\\
    & & \quad - k_2 [\ce{S2}]  &\left[\diagram{enzyme--trees-E.pdf}{0.2}\right]\,, \\
    \frac{N_{\ce{E}}}{L_{\ce{E}}} J\subsc{ext} &= & \qquad k_6 & \left[\diagram{enzyme--trees-EP.pdf}{0.2}\right] \\
    & & - k_{-6} [\ce{P}] &\left[\diagram{enzyme--trees-E.pdf}{0.2}\right] \,.
  \end{alignat*}
  Next, we multiply the remaining pseudo-first-order rate constants into the diagrams and highlight them in blue.
  This leads us to
  \begin{align*}
    \frac{N_{\ce{E}}}{L_{\ce{E}}} J\subsc{int} = \qquad & \left[\diagram{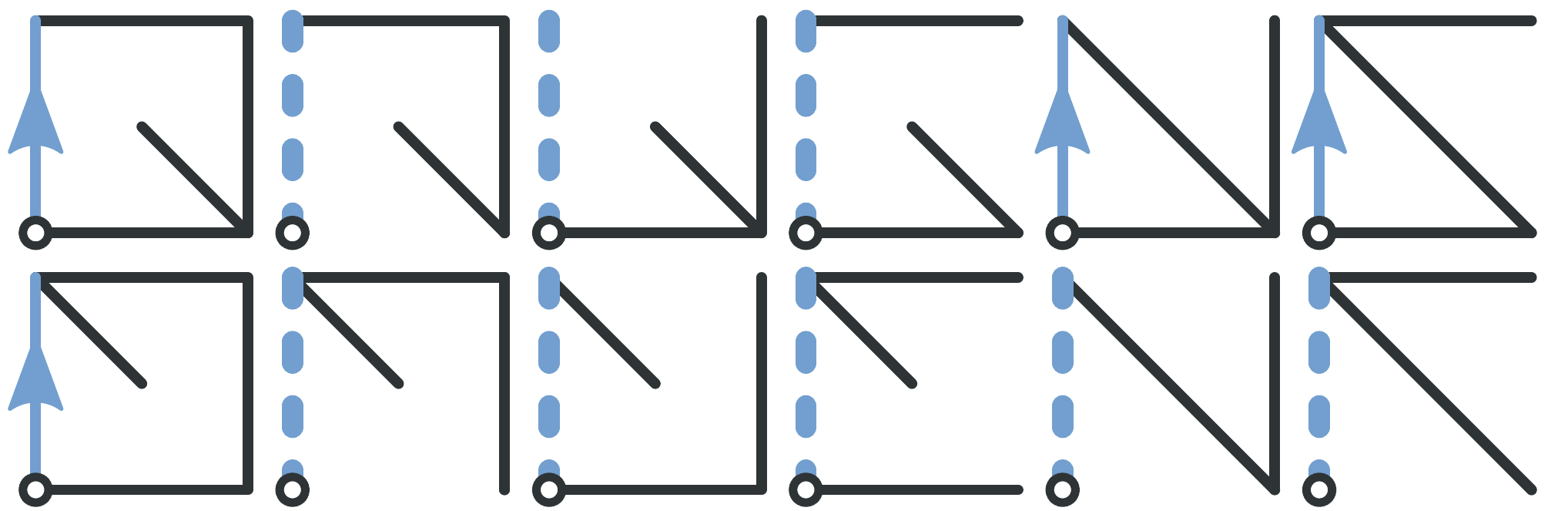}{0.2}\right]\\
    \quad - &\left[\diagram{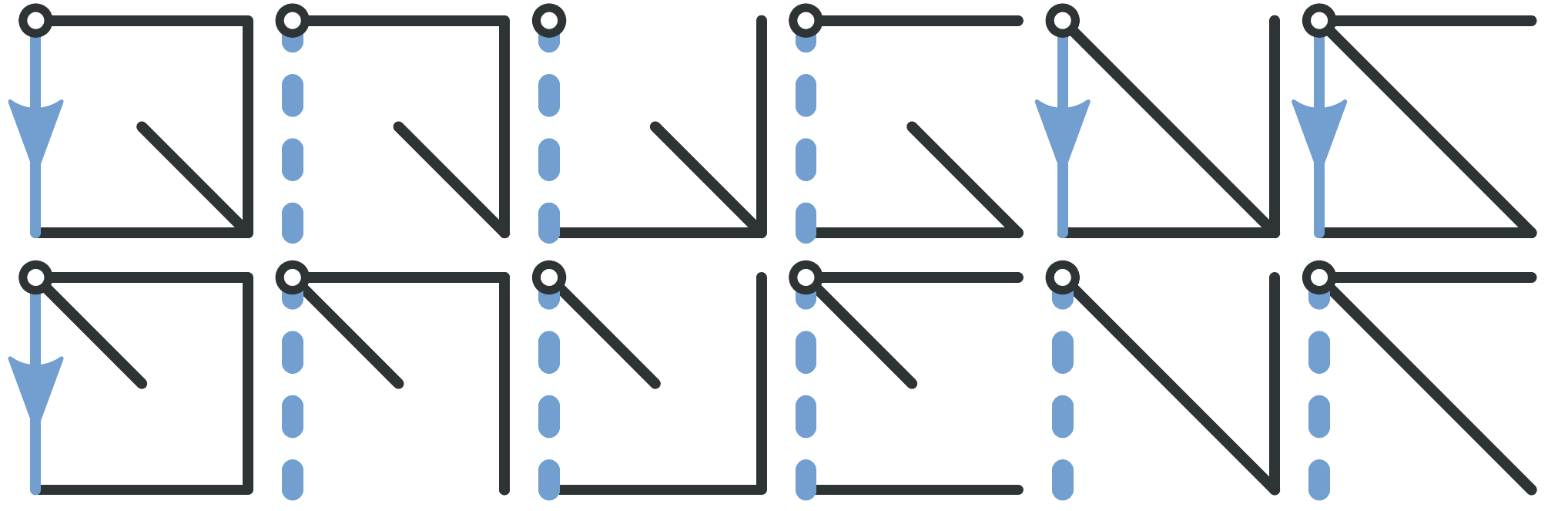}{0.2}\right]\,, \\
    \frac{N_{\ce{E}}}{L_{\ce{E}}} J\subsc{ext} = \qquad & \left[\diagram{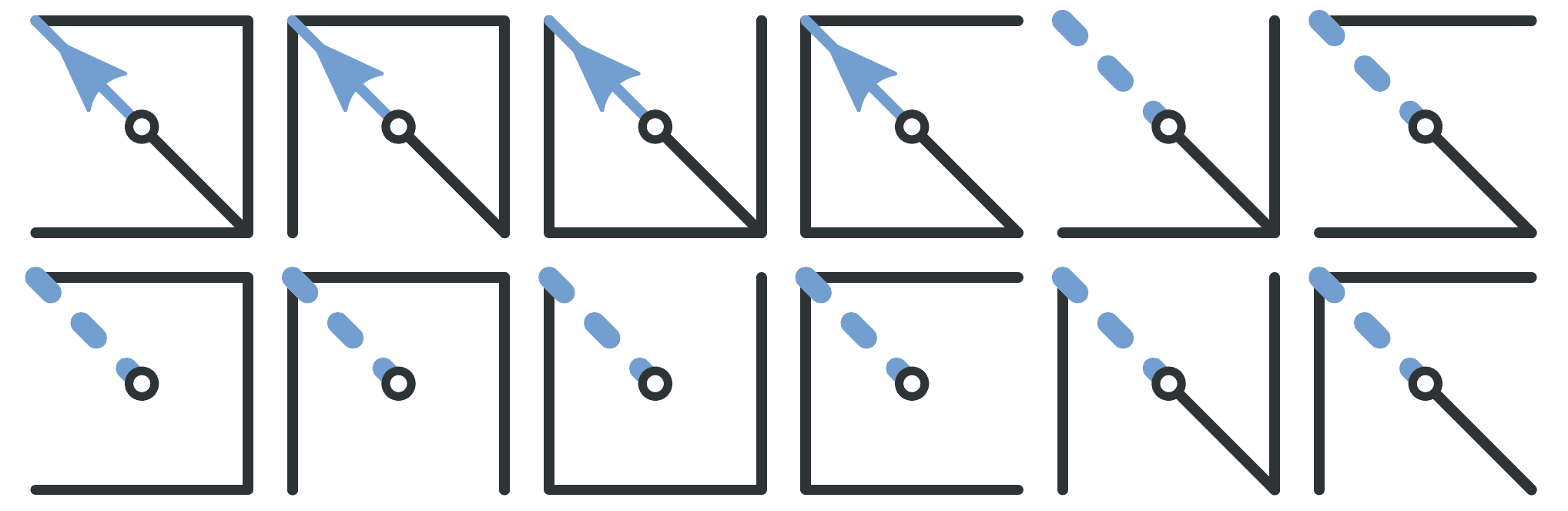}{0.2}\right] \\
    - &\left[\diagram{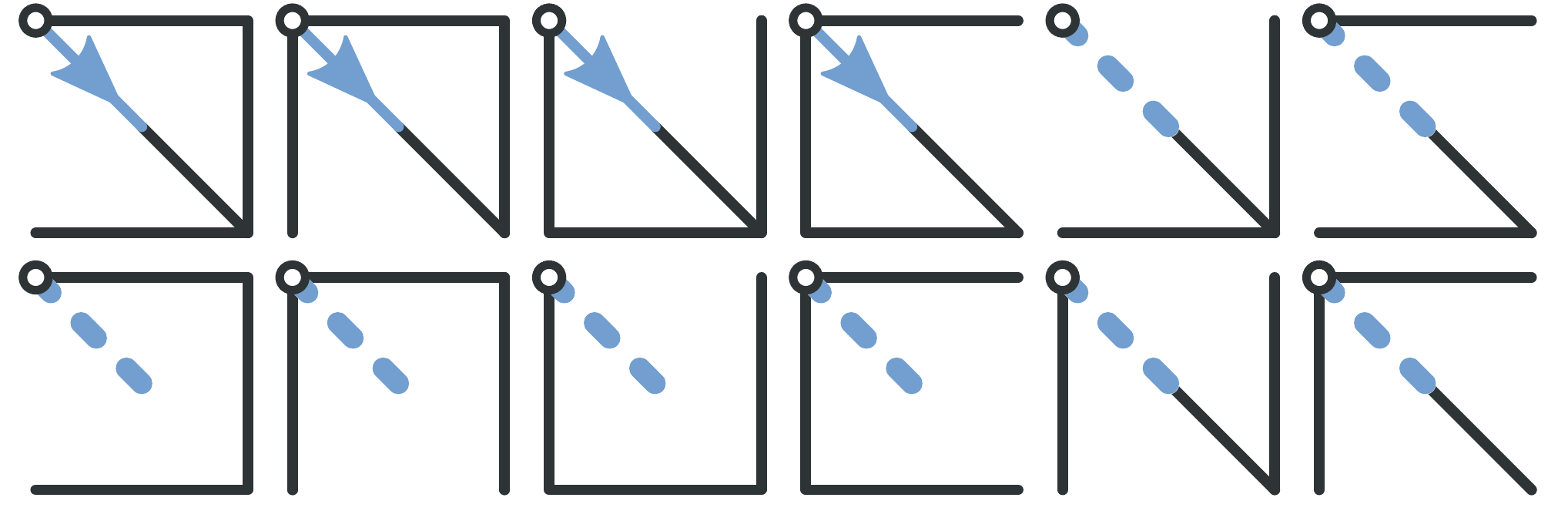}{0.2}\right] \,.
  \end{align*}
  Note how some of the diagrams did not contain that edge before, leading to a circuit in the new diagrams.
  The new pseudo-first-order rate constant carries an arrowhead to highlight the orientation of that edge.
  The black edges remain oriented along the other black edges towards the circled vertex.
  The remaining diagrams already contained the reverse pseudo-first-order rate constant for the newly incorporated edge.
  The product of these forward and backward pseudo-first-order rate constants is highlighted as a dashed blue edge without arrowhead.
  The latter tree diagrams appear on both sides of the minus signs and can be cancelled.
  Thus the currents are
  \begin{align*}
    \frac{N_{\ce{E}}}{L_{\ce{E}}} J\subsc{int} &= \left[ \diagram{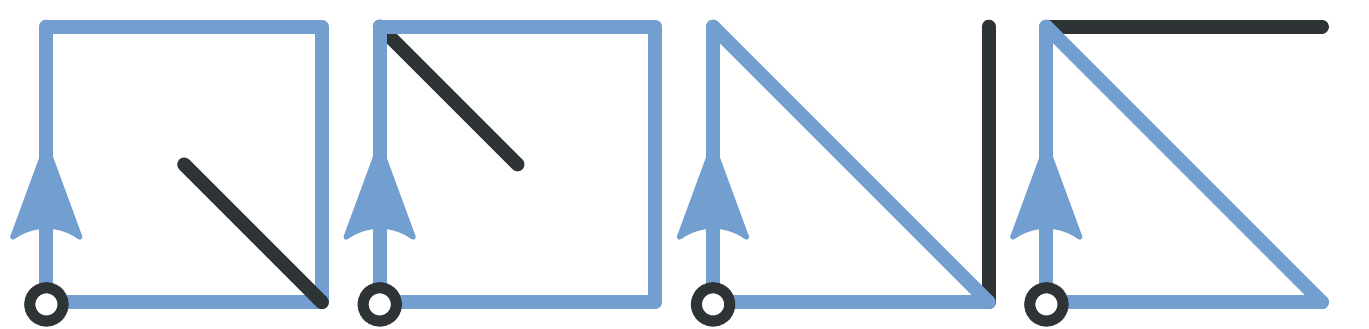}{0.2}\right]
    - \left[\diagram{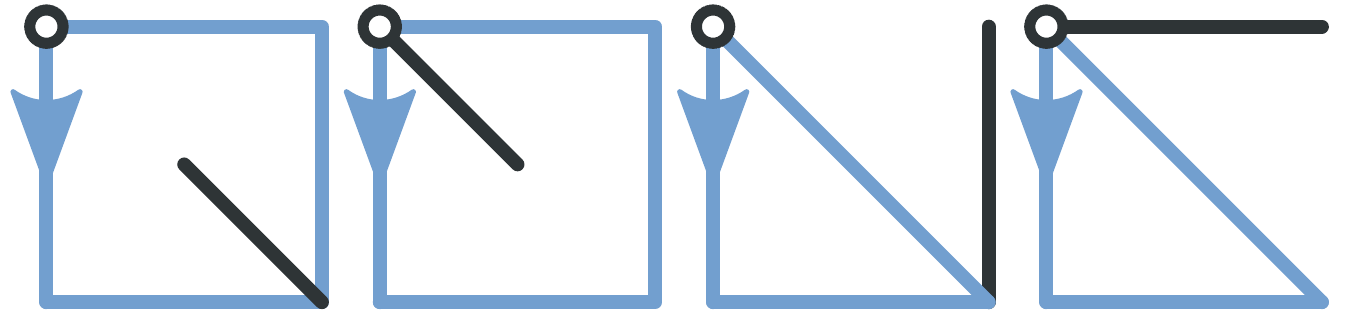}{0.2} \right] \,,\\
    \frac{N_{\ce{E}}}{L_{\ce{E}}} J\subsc{ext} &= \left[ \diagram{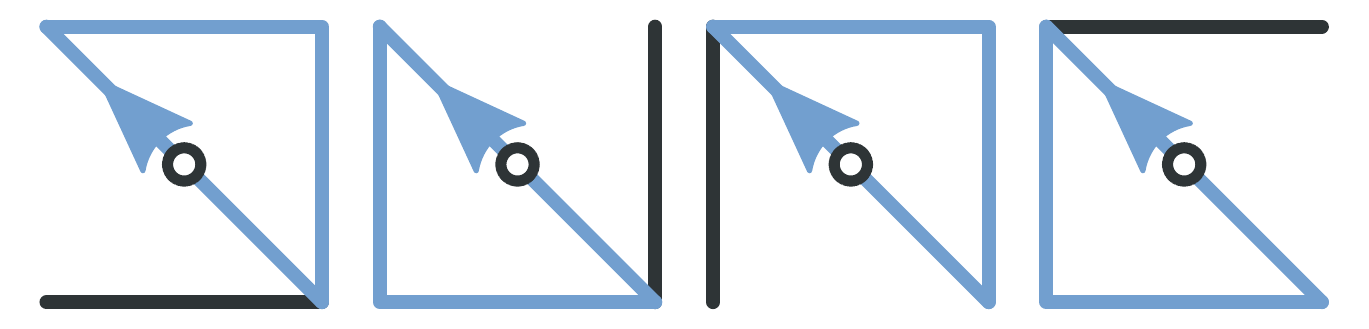}{0.2}\right]
  - \left[\diagram{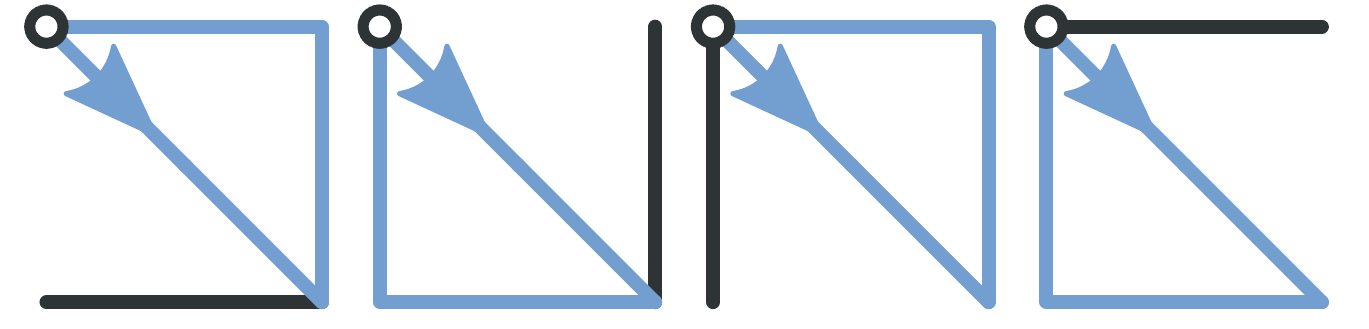}{0.2}\right] \,.
  \end{align*}
  Here, we highlight the entire circuits in blue to emphasize the common factors in the remaining terms.
  Note that the square representing the internal cycle remained in the internal cycle current on both sides of the minus sign.
  However, Wegscheider's conditions, \Eq{expl:zero-force}, ensure that these terms cancel as well.
  Furthermore, Wegscheider's conditions allow us to express the diagrams containing the lower triangle with the upper triangle:
  \begin{align*}
    \diagram{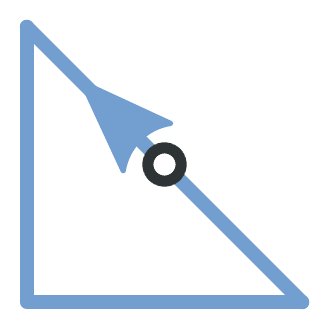}{0.2} = \diagram{enzyme--psi+_lower.pdf}{0.2}\times \frac{\diagram{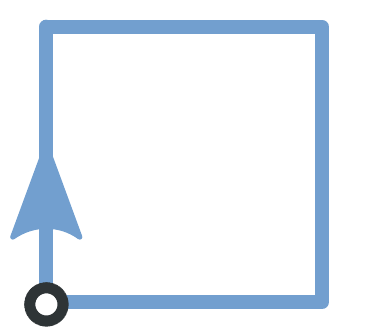}{0.2}}{\diagram{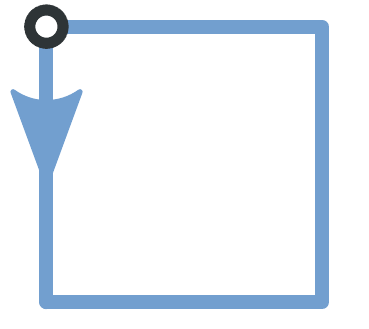}{0.2}} &= \diagram{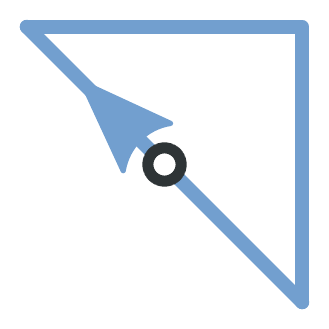}{0.2}\frac{k_{-2}k_{-3}}{k_{-1}k_{-4}}\,, \\
    \diagram{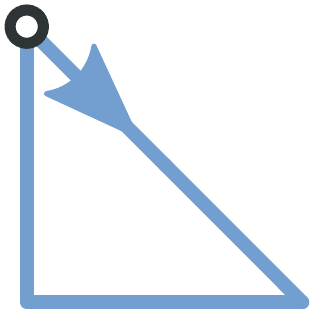}{0.2} = \diagram{enzyme--psi-_lower.pdf}{0.2}\times \frac{\diagram{enzyme--psi_int-.pdf}{0.2}}{\diagram{enzyme--psi_int+.pdf}{0.2}} &= \diagram{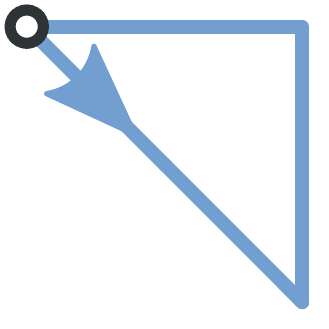}{0.2}\frac{k_{2}k_{3}}{k_{1}k_{4}} \,.
  \end{align*}

Overall, the currents expressed with rate constants and concentrations are
  \begin{align*}
    J\subsc{int}
  &= \frac{L_{\ce{E}}k_{2}k_{3}}{N_{\ce{E}}k_{1}}\! \left( \frac{k_{-1}}{k_{4}} +  [\ce{S2}] \right)\! \left(  k_{-1}k_{-4}k_{-5}k_{-6} [\ce{P}] - k_{1}k_{4}k_{5}k_{6} [\ce{S1}] [\ce{S2}] \right)\,,
  \\
  J\subsc{ext}
   &= \frac{L_{\ce{E}}}{N_{\ce{E}}} \left( k_{3}[\ce{S1}] + \frac{k_{2}k_{3}[\ce{S2}]}{k_{1}} + k_{-2} + \frac{k_{-2}k_{-3}}{k_{-4}} \right)\times \\
 & \qquad \left(  k_{1}k_{4}k_{5}k_{6} [\ce{S1}] [\ce{S2}] - k_{-1}k_{-4}k_{-5}k_{-6} [\ce{P}]\right) \,.
  \end{align*}
}

\nw{
\subsection{Kinetic rate laws for the active transporter}
\label{sec:appendix--ratelaws--transporter}
}

\nw{
We proceed analogously to the previous calculation for the enzymatic catalysis:
Plug the tree diagrams from appendix~\ref{sec:appendix--steady-state--transporter} into
\begin{align*}
  J\subsc{cat} &= J_{2} = k_{2} [\ce{S}] [\ce{HM}]  - k_{-2} [\ce{HMS}]\,, \\
  J\subsc{sl} &= -J_{1} = k_{-1} [\ce{HM}] - k_{1} [H^{+}\subsc{a}] [\ce{^{-}M}]\,,
\end{align*}
and cancel all diagrams that do not contain a circuit.
This leads us to 
\begin{align*}
  \frac{N_{\ce{M}}}{L_{\ce{M}}} J\subsc{cat} &= \left[\diagram{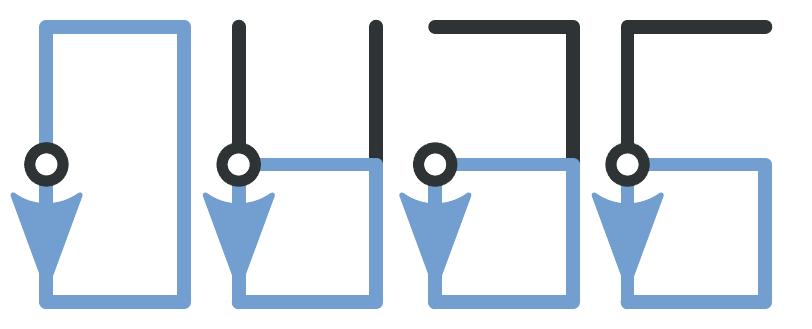}{0.2}\right] - \left[ \diagram{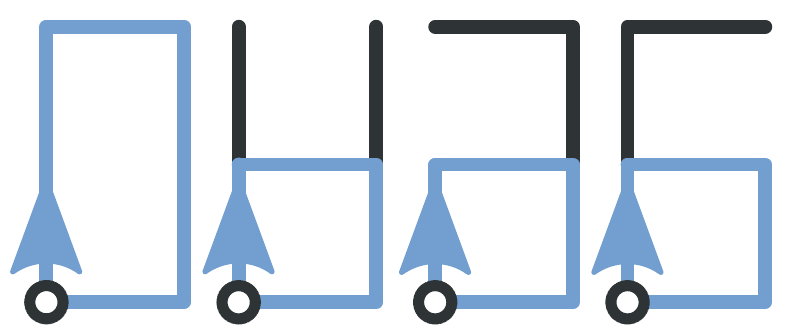}{0.2} \right]\,,\\
  \frac{N_{\ce{M}}}{L_{\ce{M}}} J\subsc{sl} &= \left[ \diagram{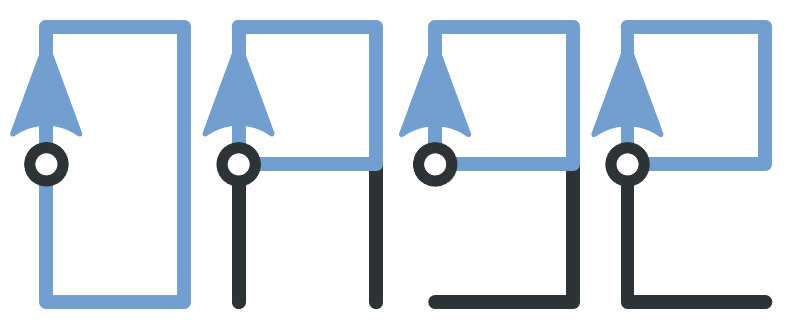}{0.2} \right] - \left[ \diagram{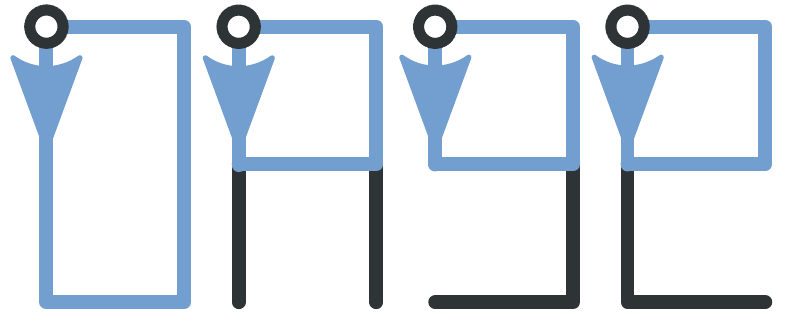}{0.2} \right]\,.
\end{align*}
Since this membrane transporter mechanism does not have an internal cycle, we cannot exploit Wegscheider's conditions to cancel more terms.
Nonetheless, we see that we can factor the circuits out of some of the terms.
Overall, we arrive at the cycle currents
\begin{align*}
  J\subsc{cat} &=: \psi^{+}\subsc{cat} - \psi^{-}\subsc{cat}\,, & 
  J\subsc{sl} &=: \psi^{+}\subsc{sl} - \psi^{-}\subsc{sl} \,.
\end{align*}
with the fluxes
\begin{align*}
  \frac{N\subsc{M}}{L_{\ce{M}}} \psi^+\subsc{cat}
  &=
  k_{1}k_{2}k_{3}k_{4}k_{5}k_{6}[\ce{H^+_a}][\ce{S}]
   + \xi\subsc{cat} k_{-7}k_{2}k_{3}k_{4}[\ce{S}]
  \,, \\
  \frac{N\subsc{M}}{L_{\ce{M}}} \psi^{-}\subsc{cat}
  &=
  k_{-1}k_{-2}k_{-3}k_{-4}k_{-5}k_{-6}[\ce{H^+_b}][\ce{P}]
  + \xi\subsc{cat} k_{7}k_{-2}k_{-3}k_{-4}[\ce{P}]
  \,,\\
  \frac{N\subsc{M}}{L_{\ce{M}}} \psi^{+}\subsc{sl}
  &=
  k_{-1}k_{-2}k_{-3}k_{-4}k_{-5}k_{-6}[\ce{H^+_b}][\ce{P}]
  +
  \xi\subsc{sl} k_{-1}k_{-5}k_{-6}k_{-7}[\ce{H^+_b}]
  \,, \\ 
  \frac{N\subsc{M}}{L_{\ce{M}}} \psi^{-}\subsc{sl}
  &=
  k_{1}k_{2}k_{3}k_{4}k_{5}k_{6}[\ce{H^+_a}][\ce{S}]
  +
  \xi\subsc{sl} k_{1}k_{5}k_{6}k_{7}[\ce{H^+_a}]
  \,,
\end{align*}
where we used the abbreviations
\begin{align*}
  \xi\subsc{cat} &:= k_{-6}k_{-5}[\ce{H^+_b}] + k_{1}k_{-5}[\ce{H^+_a}][\ce{H^+_b}] + k_{6}k_{1}[\ce{H^+_a}]\,,\\
  \xi\subsc{sl} &:=  k_{3}k_{4} + k_{-2}k_{4} + k_{-3}k_{-2}\,.
\end{align*}
}

\nw{
\subsection{Kinetic rate laws for generic catalysts}
\label{sec:appendix--fluxes--generic}
}

\new{

  \nw{By making} use of the graph theory notation introduced in appendix~\ref{sec:appendix--steady-state}\nw{, we can generalize the above calculations to generic catalysts}.

  Before proceeding with calculations, we \nw{need a general method to determine the cycle currents from individual reaction currents.
  To that end, we} construct a special spanning tree \(\tau^{*}\) for the graph \(\mathcal{G}\) of the open system:
(1) We start with the \emph{closed} system and determine its internal cycles \(\ker \stoich\).
We take the set \(\mathcal{I}\subset\mathcal{R}\) of edges that the internal cycles are supported on.
(2) Consider this set of edges \(\mathcal{I}\subset\mathcal{G}\) as a subgraph of the \emph{open} network.  Choose a spanning tree \(\tau_\mathcal{I}\) for this subgraph.
(3) Complete \(\tau_\mathcal{I}\) to a spanning tree \(\tau^*\) of \(\mathcal{G}\).
All the edges not contained in the spanning tree are the \emph{chords}.

There is a special connection between chords and circuits first highlighted by Schnakenberg~\cite{Schnakenberg1976}:
The spanning tree alone, by definition, does not contain any circuit.
Adding a chord to the spanning tree gives rise to a circuit composed of the chord together with edges from the spanning tree.
Furthermore, by construction every chord gives rise to a \emph{different} circuit and the set of these circuits form a basis of the cycle space \(\ker \stoich^{X}\).
In this context the circuits associated to chords are also called \emph{fundamental cycles}.
The currents on the chords then are identical to the steady-state currents along the fundamental cycles of the chords~\cite{Schnakenberg1976}.

The special spanning tree \(\tau^{*}\) that we constructed is separating the chords into two sets:
Each chord in \(\mathcal{I}\) gives rise to an \emph{internal cycle},
while the chords not in \(\mathcal{I}\) give rise to the emergent cycles.
This construction provides a basis for the entire cycle space, yet keeps the internal cycles and the emergent cycles separated.
Therefore we call it a \emph{separating spanning tree}.
}

\nw{It is worth noting that not every basis of circuits can be expressed as fundamental cycles of a spanning tree.
This technical detail, however, has no impact on our results.
Different bases are just different representations of the same space.
In the following we assume a spanning tree mainly for convenience.}

Let \(j\to i\) be the chord of an emergent cycle.
Then the current through that chord is
\begin{align*}
  J_{ij} &= \tilde{k}_{ij}(\vec{y}) x_{j} - \tilde{k}_{ji}(\vec{y}) x_{i}\\
  &= \frac{L}{N(\vec{y})} \left[ \tilde{k}_{ij}(\vec{y}) \sum_{\tau \in \mathcal{T}_j} \prod_{\rho \in \tau} \tilde{k}_{\rho}(\vec{y}) - \tilde{k}_{ji}(\vec{y}) \sum_{\tau \in \mathcal{T}_i} \prod_{\rho \in \tau} \tilde{k}_{\rho}(\vec{y}) \right] \,.
\end{align*}
Next, we note that a lot of terms cancel by taking this difference.
All the spanning trees that contain the edge \(i\to j\) or \(j\to i\), respectively, appear with both plus and minus sign:
\begin{align*}
  J_{ij} &= \frac{L}{N(\vec{y})} \underbrace{\left[ \tilde{k}_{ij}(\vec{y}) \sum_{\substack{\tau \in \mathcal{T}_j\\i\to j \in \tau}} \prod_{\rho \in \tau} \tilde{k}_{\rho}(\vec{y}) - \tilde{k}_{ji}(\vec{y}) \sum_{\substack{\tau \in \mathcal{T}_i\\j\to i \in \tau}} \prod_{\rho \in \tau} \tilde{k}_{\rho}(\vec{y}) \right]}_{=0}\\
  &\quad + \frac{L}{N(\vec{y})} \left[ \tilde{k}_{ij}(\vec{y}) \sum_{\substack{\tau \in \mathcal{T}_j\\i\to j \notin \tau}} \prod_{\rho \in \tau} \tilde{k}_{\rho}(\vec{y}) - \tilde{k}_{ji}(\vec{y}) \sum_{\substack{\tau \in \mathcal{T}_i\\j \to i \notin \tau}} \prod_{\rho \in \tau} \tilde{k}_{\rho}(\vec{y}) \right] \,.
\end{align*}
After cancelling these spanning tree contributions, we define the \emph{apparent cycle fluxes} as 
\begin{align}
  \psi_{ij} :=\frac{L}{N(\vec{y})} \tilde{k}_{ij}(\vec{y}) \sum_{\substack{\tau \in \mathcal{T}_j\\i\to j \notin \tau}} \prod_{\rho \in \tau} \tilde{k}_{\rho}(\vec{y}) \,. \label{eq:coarse-grained-rates}
\end{align}
We obviously have \(J_{ij} = \psi_{ij} - \psi_{ji}\).
Thus the apparent cycle fluxes serve as kinetic rate laws for the coarse-grained reactions.

There is, technically speaking, no strict necessity to cancel the spanning tree contributions in order to arrive at expressions that can serve as coarse-grained kinetic rate laws.
Keeping the spanning tree contributions results in the apparent fluxes of the substrates/products that are being produced/consumed along the chord.
This is a natural choice for dealing with data from isotope labelling experiments.
With this definition for \new{kinetic rate laws}, however, the flux--force relation is not satisfied -- even in the case of a single emergent cycle~\cite{Wiechert2007}.
In contrast, our definition of apparent fluxes resembles the apparent \emph{cycle} fluxes, rather than apparent exchange fluxes.
Comparing the apparent cycle fluxes with the net force along the emergent cycle, we do have a flux--force relation, as shown in 
the next section.


\section{Proof of the flux--force relation}
\label{sec:appendix--flux-force}

Before we prove the flux--force relation, we rewrite the apparent fluxes for the emergent cycles derived in \Eq{coarse-grained-rates}.
This simplifies the final proof considerably.
To that end, we observe that adding a chord to a spanning tree not containing this chord always creates a \new{circuit}.
Since in \Eq{coarse-grained-rates} we sum over all possible spanning trees, the same circuits re-appear in several summands.
We now re-sort the sums to first run over distinct circuits, and then sum over the remainders of the spanning trees.
For that we need some notation.

For any circuit \(c\) we abbreviate the product of pseudo-first-order rate constants along it as \( w(c) = \prod_{\rho\in c} \tilde{k}_{\rho}(\vec{y}) \)\,.
The net force along a circuit thus is concisely written as
\begin{align}
  -\Delta_{c} G = RT \sum_{\rho\in c} \ln \frac{\tilde{k}_{\rho}(\vec{y})}{\tilde{k}_{-\rho}(\vec{y})} = RT \ln \frac{w(c)}{w(-c)}\,. \label{eq:coarse-grained-net-force}
\end{align}
\new{Here, \(-c\) refers to traversing the circuit \(c\) with reversed orientation.}
For any circuit, \(c\), we furthermore define \(\mathcal{F}(c)\) to be the set of subforests of \(\mathcal{G}\) \new{
that (i) do not contain any edge of \(c\), (ii) span the rest of the graph, and (iii) are directed towards the circuit \(c\).}
Analogously to the product of rate constants along a circuit, for this set of subforests we denote the sum of products of rate constants as
\begin{align*}
  \xi(c) \coloneqq \sum_{f\in \mathcal{F}(c)} \prod_{\rho\in f} \tilde{k}_{\rho}(\vec{y})\,.
\end{align*}
By construction, \(\xi(c) = \xi(-c)\) since the set \(\mathcal{F}(c)\) does not depend on the orientation of \(c\).
\new{Let \(\mathcal{C}_{ij}\) be the set of circuits traversing the edge \(j\to i\).
Note that these circuits are exactly the ones appearing in \Eq{coarse-grained-rates}
.}

With this notation we rewrite the apparent cycle fluxes in the following way:
\begin{align*}
  \psi_{ij} = \frac{L}{N(\vec{y})} \sum_{c\in \mathcal{C}_{ij}} w(c) \xi(c)\,.
\end{align*}
This rewriting is not limited to the case of a single emergent cycle.
\new{In fact, we used this form to express the apparent cycle fluxes of the active membrane transporter in \nw{appendix~\ref{sec:appendix--ratelaws--transporter}}.}

We now prove the flux--force relation -- under the assumption that there is exactly one emergent cycle \(c_{\eta}\) with chord \(\eta = j\to i\).
Let \(-\Delta_{\eta} G\) be the net force along this cycle and let \(J_{\eta}\) be its current at steady state.
\new{Let furthermore \(\tau^{*}\) be a separating spanning tree, as we defined in appendix~\ref{sec:appendix--fluxes--generic}.}

Having only one emergent cycle means that for every circuit \(c\in \mathcal{C}_{ij}\) we have one of the following cases:
\begin{itemize}
  \item The circuit is formed by following the \new{separating} spanning tree \new{from vertex \(i\) back to \(j\)}, in which case it is exactly the emergent cycle: \(c=c_{\eta}\).
  \item The circuit is formed by traversing more chords, in which case it can be written as \(c= c_\eta + \gamma\) where \(\gamma\in \ker \stoich\) is an internal cycle.
    In this case we have \(\frac{w(c)}{w(-c)}= \frac{w(\gamma)}{w(-\gamma)}\frac{w(c_\eta)}{w(-c_\eta)}=\frac{w(c_\eta)}{w(-c_\eta)}\) due to Wegscheider's conditions.
\end{itemize}
In any case we can write \(w(\pm c) = \zeta(c) w(\pm c_\eta)\) where \(\zeta(c)=\zeta(-c)\) is a symmetric factor.
Overall, the apparent fluxes for the emergent cycle are
\begin{align*}
  \psi_{ij} &= \frac{L}{N(\vec{y})} \sum_{c\in \mathcal{C}_{ij}} w(c) \xi(c)
  = \frac{L}{N(\vec{y})} \left[ \sum_{c\in \mathcal{C}_{ij}} \xi(c) \zeta(c)\right]w(c_\eta) \,.
\end{align*}
By construction, \(\xi\) and \(\zeta\) are symmetric and also any sum over these terms is symmetric.
Consequently, the apparent forward and backward fluxes of the emergent cycle satisfy
\begin{align*}
  \frac{\psi_{ij}}{\psi_{ji}} = \frac{\frac{L}{N(\vec{y})} \left[ \sum_{c\in \mathcal{C}_{ij}} \xi(c) \zeta(c)\right]w(c_\eta) 
}{\frac{L}{N(\vec{y})} \left[ \sum_{c\in \mathcal{C}_{ij}} \xi(\new{-c}) \zeta(\new{-c})\right]w(-c_\eta) 
}  = \frac{w(c_\eta)}{w(-c_\eta)}
\end{align*}
which, together with \Eq{coarse-grained-net-force}, concludes the proof.

\new{
From this proof it is evident, why the flux--force relation breaks down once there are several emergent cycles with non-zero forces:
In the case where a circuit \(c\in\mathcal{C}_{ij}\) is not identical to the emergent cycle \(c_{\eta}\), we can still write it as \(c = c_{\eta} + \gamma\).
However, now \(\gamma\) need not be an internal but might be another emergent cycle.
Therefore, Wegscheider's condition does not apply to it, thus \(w(\gamma)\) and hence \(\zeta(c)\) need not be symmetric.
As a consequence, the ratio of apparent forward and backward cycle fluxes cannot be expressed by the force of the emergent cycle \(-\Delta_{\eta}G\) alone.
}

\nw{The proof also shows why the choice of a separating spanning tree is mainly for convenience.
In the case of a single emergent cycle, the exact basis for the internal cycles does not matter and you can always find an appropriate separating spanning tree.
In the case of several emergent cycles, there is no simple and direct relation between the force and the fluxes of a cycle.
The only consistency requirement is the entropy-production rate.
However, the entropy-production rate is a scalar and thus invariant under change of basis.
Furthermore, it involves only the forces and the currents of the cycles.
This imposes no restrictions on the individual forward and backward fluxes.}

\bibliography{enzymes}


\end{document}